\documentclass[reprint, twocolumn, aps, prl, longbibliography, superscriptaddress, showpacs, preprintnumbers]{revtex4-2}
\usepackage{graphicx}%
\usepackage{siunitx}
\usepackage{bm}
\usepackage[caption=false]{subfig}
\usepackage{mathtools}
\usepackage{multirow}
\usepackage{amssymb}
\usepackage{braket}
\usepackage{amsmath}
\usepackage{placeins}
\usepackage[dvipsnames]{xcolor}
\usepackage{spreadtab}
\usepackage{dsfont}
\usepackage[colorlinks=true,
            linkcolor=blue,
	        citecolor=blue,
	         urlcolor=blue]{hyperref}

\usepackage{tikz}
\usetikzlibrary{matrix,decorations.pathreplacing,calc,fit,backgrounds}

\pgfmathsetmacro{\myscale}{2}
\pgfkeys{tikz/mymatrixenv/.style={decoration={},every left delimiter/.style={xshift=8pt},every right delimiter/.style={xshift=-8pt}}}
\pgfkeys{tikz/mymatrix/.style={matrix of math nodes,nodes in empty cells,
left delimiter={[},right delimiter={]},inner sep=1pt,outer sep=1.5pt,
column sep=8pt,row sep=8pt,nodes={minimum width=20pt,minimum height=10pt,
anchor=center,inner sep=0pt,outer sep=0pt,scale=\myscale,transform shape}}}
\pgfkeys{tikz/mymatrixbrace/.style={decorate,thick}}

\tikzset{greenish/.style={
    fill=green!50!lime!60,draw opacity=0.7,
    draw=green!50!lime!60,fill opacity=0.1,
  },
  cyanish/.style={
    fill=cyan!90!blue!60, draw opacity=0.4,
    draw=blue!70!cyan!30,fill opacity=0.1,
  },
  orangeish/.style={
    fill=orange!90, draw opacity=0.8,
    draw=orange!90, fill opacity=0.1,
  },
  brownish/.style={
    fill=brown!70!orange!40, draw opacity=0.4,
    draw=brown, fill opacity=0.3,
  },
  purpleish/.style={
    fill=violet!90!pink!20, draw opacity=0.5,
    draw=violet, fill opacity=0.3,    
  }}

\newcommand\bk{{\mathbf{k}}}
\newcommand\bq{{\mathbf{q}}}
\newcommand\br{{\mathbf{r}}}
\newcommand\bR{{\mathbf{R}}}
\newcommand{\nk}[0]{{n\bk}}

\newcommand{\qka}[0]{{\bq \kappa\alpha}}

\newcommand{\qkpb}[0]{{\bq \kappa'\beta}}

\newcommand{\Fig}[1]{Fig.~\ref{#1}}

\newcommand{\Figu}[1]{Figure~\ref{#1}}
\newcommand{\Eq}[1]{Eq.~\eqref{#1}}
\newcommand{\Eqs}[1]{Eqs.~\eqref{#1}}

\DeclarePairedDelimiterX\abs[1]{\lvert}{\rvert}{#1}

\begin{document}

\title{Verification and Validation of zero-point electron-phonon renormalization of the bandgap, mass enhancement, and spectral functions}

\author{Samuel Ponc\'e}
\email{samuel.ponce@uclouvain.be}
\affiliation{%
European Theoretical Spectroscopy Facility and Institute of Condensed Matter and Nanosciences, Université catholique de Louvain, Chemin des Étoiles 8, B-1348 Louvain-la-Neuve, Belgium. 	
}%
\affiliation{%
WEL Research Institute, avenue Pasteur 6, 1300 Wavre, Belgium.		
}%
\author{Jae-Mo Lihm}
\affiliation{%
European Theoretical Spectroscopy Facility and Institute of Condensed Matter and Nanosciences, Université catholique de Louvain, Chemin des Étoiles 8, B-1348 Louvain-la-Neuve, Belgium. 	
}%
\affiliation{%
Department of Physics and Astronomy, Seoul National University, Seoul 08826, Korea;\\
Center for Correlated Electron Systems, Institute for Basic Science, Seoul 08826, Korea;\\
and Center for Theoretical Physics, Seoul National University, Seoul 08826, Korea
}%
\author{Cheol-Hwan Park}
\affiliation{%
Department of Physics and Astronomy, Seoul National University, Seoul 08826, Korea;\\
Center for Correlated Electron Systems, Institute for Basic Science, Seoul 08826, Korea;\\
and Center for Theoretical Physics, Seoul National University, Seoul 08826, Korea
}%

\date{\today}

\begin{abstract}
Verification and validation of methods and first-principles software are at the core of computational solid-state physics but are too rarely addressed. 
We compare four first-principles codes: \textsc{Abinit}, \textsc{Quantum ESPRESSO}, \textsc{EPW}, \textsc{ZG}, and three methods: (i) the Allen-Heine-Cardona theory using density functional perturbation theory (DFPT), (ii)  the Allen-Heine-Cardona theory using Wannier function perturbation theory (WFPT), and (iii) an adiabatic non-perturbative frozen-phonon method. 
For these cases, we compute the real and imaginary parts of the electron-phonon self-energy in diamond and BAs, including dipoles and quadrupoles when interpolating.  
We find excellent agreement between software that implements the same formalism as well as good agreement between the DFPT and WFPT methods.
Importantly, we find that the Deybe-Waller term is momentum dependent which impacts the mass enhancement, yielding approximate results when using the Luttinger approximations. 
Finally, we compare the electron-phonon spectral functions between \textsc{Abinit} and \textsc{EPW} and find excellent agreement even away from the band edges.       
\end{abstract}

\maketitle

\section*{INTRODUCTION}

The increasing difficulty in reproducing psychological experiments led to a \emph{replication crisis}~\cite{Pashler2012}, affecting all areas of science.  
The field of first-principles calculations of materials properties has not been spared given the growing complexity of software implementation, supporting heterogeneous architecture and multiple types of parallelization.   
As a result, many frameworks~\cite{Yuan2010}, initiatives~\cite{Korth2009} and global efforts~\cite{Huber2021} have been started in the last decade which 
culminated in two large cross-validation studies of the equation of states of elemental crystals~\cite{Lejaeghere2016} as well as oxides~\cite{Bosoni2023}.




However, most of these verification efforts have been concentrated on basic ground-state properties based on density functional theory (DFT).
Advanced features such as excited state properties have comparatively received less attention~\cite{Setten2015,Setten2017,Nguyen2018,Govoni2018,Alvertis2024}.   
In particular, the validation of first-principles methods to compute the electronic bandstructure renormalization due to electron-phonon coupling has only been investigated in the following list of works, that we hope is exhaustive. 
Reference~\onlinecite{Ponce2015} studies a comparison between adiabatic versus non-adiabatic versions of the Allen-Heine-Cardona (AHC) theory~\cite{Allen1976,Allen1981}, ref.~\onlinecite{Miglio2020} compares perturbation methods based on the density functional perturbation theory (DFPT) versus adiabatic supercell methods, ref.~\onlinecite{Engel2022} contrasts norm-conserving pseudopotentials and the projector augmented wave (PAW) method~\cite{Blochl1994a} with long-range Fr\"ohlich corrections, ref.~\onlinecite{Zacharias2020a} assess harmonic versus anharmonic using \emph{ab-initio} molecular dynamics (MD), ref.~\onlinecite{Lihm2020} performed a comparison between the AHC method with and without bands off-diagonal elements in the electron-phonon self-energy, and ref.~\onlinecite{Yang2021} against many-body methods with non-adiabatic effects. 
There are nonetheless still many open questions.

In addition, the \emph{verification} of codes, meaning the verification that independent implementation of the \emph{same} method gives the 
same result, is almost non-existent despite being hugely important. 
To the authors' knowledge, one verification of the zero-point renormalization (ZPR) between the \textsc{Abinit}~\cite{Gonze2016,Gonze2020}
and the \textsc{Quantum ESPRESSO}~\cite{Giannozzi2017} plus \textsc{Yambo}~\cite{Sangalli2019} has been performed in Ref.~\onlinecite{Ponce2014}, a verification of the ZPR between \textsc{FHI-aims}~\cite{Blum2009} and \textsc{Abinit} has been done in Ref.~\onlinecite{Shang2021}, and between \textsc{VASP}~\cite{Kresse1993} and \textsc{Abinit} in Ref.~\onlinecite{Engel2022}.

In this work, we \emph{verify} the adiabatic and non-adiabatic~\cite{Ponce2015,Miglio2020} AHC implementation in \textsc{Abinit}~\cite{Gonze2011,Gonze2020} and \textsc{Quantum ESPRESSO}~\cite{Lihm2020} by computing the ZPR, mass-enhancement, and spectral functions of the infrared-inactive diamond and infrared-active BAs materials.
We also \emph{validate} the accuracy of alternative approaches based on Wannier function perturbation theory (WFPT)~\cite{Lihm2021} as implemented in \textsc{EPW}~\cite{Giustino2007,Ponce2016a,Lee2023}, various approximations inspired by the Luttinger's theorem~\cite{Luttinger1960b}, and the adiabatic non-perturbative Zacharias and Giustino (ZG)~\cite{Zacharias2020,Lee2023} approach based the special displacement method~\cite{Zacharias2016}.
To our knowledge, we note that among the free open source software allowing first-principles calculations of electron-phonon coupling, \textsc{Perturbo}~\cite{Zhou2021}, \textsc{EPIq}~\cite{Marini2024}, and \textsc{Phoebe}~\cite{Cepellotti2022} do not currently have the functionality of computing the Debye-Waller self-energy. 
Therefore, we have not investigated them in our study.

We first briefly present the theory of the different approaches, including their approximation and how each method is expected to connect with the others. 
To achieve this, we delve into the relevant technical details and present numerical results for the case of diamond and BAs. 
We stress that such verification and validation efforts are crucial and that, in the context of this work, we have fixed some bugs, implemented missing features, and improved the documentation of the codes. 
The details of these improvements can be found in the commit history of the public respective codes and are summarized in the Supplemental Information (SI)~\cite{SI}.

Then, we discuss the specific case of the mass-enhancement due to electron-phonon coupling. 
In this case, the electron momentum dependence of the electron-phonon self-energy (and in particular that of the Debye-Waller (DW) term) becomes important.
However, in practice, the momentum dependence or even the entire DW term is often neglected~\cite{Ponce2016a,Verdi2017,Abramovitch2023}. 
In passing we note that the DW factor, $W=e^{\frac{-1}{3}\Delta k^2 \langle U^2(T)\rangle}$, used to describe the attenuation of scattering due to thermal motion~\cite{Debye1913,Waller1923}, should not be confused with the DW self-energy described here and used to compute the zero-point renormalization in the AHC theory~\cite{Allen1976,Allen1981}. 
In both cases, the mean-squared vibrational displacement $\langle U^2(T)\rangle$ is used, but in the case of the DW self-energy, the electron-phonon coupling (EPC) matrix elements are also involved. 
Interestingly, the DW factor was shown to be non-dispersive in aluminum~\cite{Zhou2020}. 
However, as we show in this manuscript, the EPC gives a momentum dependence to the DW self-energy and we discover that such dependence can be as large as 10\%, questioning the momentum independence approximation. 
Another important finding of this work is the understanding why in diamond the ZPR of the conduction band minimum converges much faster than the ZPR of the valence band maximum when using a sum-over-state approach. 
We also clarified why spin-orbit coupling is important for mobility~\cite{Ponce2018} but not for ZPR~\cite{BrousseauCouture2023} even though they are linked by a Kramers-Kronig relation.  
Finally, we show the results for the electronic spectral function due to electron-phonon coupling. 
Such quantity adds a frequency dependence and is therefore more sensitive. 
For example, we note that perfect fulfillment of particle conservation when integrating the spectral function on the frequency axis up to the Fermi level is challenging and requires a large frequency range. 
Nonetheless, beyond small localized differences, we find excellent agreement on the spectral function between the \textsc{Abinit} and \textsc{EPW} code. 

\section*{RESULTS}
\subsection{Electronic bandstructure renormalization due to electron-phonon coupling} \label{sec:theory}

The energy and momentum dispersion of electrons in bulk solids, the one-particle spectral function $A_{n\bk}(\omega)$, can be measured accurately by angle-resolved photoemission spectroscopy (ARPES), as long as surface effects~\cite{Krempask2010} and photon-electron matrix elements~\cite{Damascelli2003} are correctly factored out. 
The electronic spectral function can have a complex structure in frequency, often consisting of a main quasiparticle peak with a certain broadening and possible satellites.
In this work, we mostly do not consider satellites and focus on the renormalization of the main peak using the Dyson-Migdal approximation~\cite{Antonius2015,Giannozzi2017,Nery2018,Lihm2020}.
The renormalized quasiparticle eigenvalues $E_{n\bk}$ are obtained by diagonalizing the renormalized Hamiltonian
\begin{equation}\label{eq:dmg}
    H_{nn'\bk}(T) = \varepsilon_{n\bk}\delta_{nn'}  + \Sigma_{nn'\bk}^{\rm H}(E_{n\bk}, E_{n'\bk}, T),
\end{equation}  
where $n$ and $n'$ label bare electron eigenstates, $\bk$ is the electron momentum, 
$\varepsilon_{n\bk}$ is the electronic eigenenergy which accounts for electron-electron interaction in a mean-field way but neglects electron-nuclei interaction,
$T$ is the temperature, and $\Sigma_{nn'}^{\rm H}$ is the Hermitian approximation~\cite{vanSchilfgaarde2006,Lihm2020} defined as
\begin{multline}
\Sigma_{nn'\bk}^{\rm H}(E_{n\bk}, E_{n'\bk}, T) \equiv \frac{1}{4}\bigl[ \Sigma_{nn'\bk}(E_{n\bk},T) + \Sigma_{nn'\bk}^\dagger(E_{n\bk}, T) \\
+ \Sigma_{nn'\bk}(E_{n'\bk}, T) + \Sigma_{nn'\bk}^\dagger(E_{n'\bk}, T) \bigr],
\end{multline}
where $\Sigma_{nn'\bk}$ is the frequency-dependent electron-phonon self-energy.

By assuming that the off-diagonal terms of $\Sigma_{nn'\bk}$ are negligible, which is valid for large bandgap semiconductors, we obtain
the usual formula
\begin{equation}\label{eq:dm}
E_{n\bk}(T) = \varepsilon_{n\bk} + \Re \Sigma_{nn\bk}(E_{n\bk}, T).
\end{equation}  
If the quasiparticle energies are close to the bare one, we can perform a Taylor expansion of $\Sigma_{nn\bk}(E, T)$ around $E=\varepsilon_{n\bk}$ and evaluate it at $E=E_{n\bk}(T)$ to 
obtain the linear approximation:
\begin{align}\label{eq:l}
E_{n\bk}^{\rm Z}(T) =& \varepsilon_{n\bk}+ Z_{n\bk}(T)\Re \Sigma_{nn\bk}(\varepsilon_{n\bk},T), \\
 Z_{n\bk}(T) \equiv & \biggl[ 1-\Re \frac{\partial \Sigma_{nn\bk}(E, T) }{\partial E}\Big|_{E=\varepsilon_{n\bk}} \biggr]^{-1}. \label{eq:lz}
\end{align}

Finally, one can approximate $Z_{n\bk}=1$ and such approximation is often called \textit{on-the-mass-shell} or the \textit{Rayleigh-Schr\"odinger} (RS) approximation:
\begin{equation}\label{eq:rs}
E_{n\bk}^{\rm RS}(T) = \varepsilon_{n\bk} + \Re \Sigma_{nn\bk}(\varepsilon_{n\bk}, T).
\end{equation} 
For clarity, we present Eqs.~\eqref{eq:dm}-\eqref{eq:rs} in the diagonal approximation but they can be solved via diagonalization as done in Eq.~\eqref{eq:dmg}.
In the case of semiconductors with large ($>0.5$~eV) bandgaps, such as the ones studied in this manuscript, the diagonal approximation is excellent~\cite{Lihm2020}.

From this temperature-dependence of the bandstructure, the temperature-dependent renormalization is defined as the difference between the bare and renormalized bands:
\begin{equation}
    \Delta E_{n\bk}(T) \equiv E_{n\bk}(T) - \varepsilon_{n\bk}.
\end{equation}
The zero-point renormalization (ZPR) is the renormalization at 0~K:
\begin{equation}
    \Delta E_{n\bk}(0) \equiv E_{n\bk}(0) - \varepsilon_{n\bk}.
\end{equation}

\subsection{Fan and Debye-Waller self-energies}
The electron self-energy due to electron-phonon interaction can be, in the lowest order, decomposed into Fan and static Debye-Waller (DW) self-energies $\Sigma_{nn'\bk}(E, T) = \Sigma_{nn'\bk}^{\rm Fan}(E, T) + \Sigma_{nn'\bk}^{\rm DW}(T)$~\cite{Schlter1975,Giustino2010,Gonze2011,Cannuccia2011,Ponce2014,Ponce2015,Giustino2017,Miglio2020,Lihm2020}.
The non-adiabatic and dynamical Fan self-energy is
\begin{multline}\label{eq:fan}
\Sigma_{nn'\bk}^{\rm Fan}(E, T) = \frac{1}{N_q} \sum_{\substack{\mathbf{q}\nu m \\ \kappa\alpha\kappa'\beta}} \frac{1}{2 \omega_{\bq\nu}} g_{mn\kappa\alpha}^*(\bk,\bq)\\
\times g_{mn'\kappa'\beta}(\bk,\bq) e_{\kappa\alpha\nu}^*(\bq)e_{\kappa'\beta\nu}(\bq)  \\
\times \sum_{\pm} \frac{n_{\bq\nu}(T) + \big[ 1 \pm (2f_{m\bk+\bq}(T) -1) \big]/2}{E-\varepsilon_{m\bk+\bq} \pm \omega_{\bq\nu} + i\eta},
\end{multline}
where $\omega_{\bq\nu}$ is the energy of a phonon with wavevector $\bq$ and mode index $\nu$, $n_{\bq\nu}(T)$ the Bose-Einstein distribution, $f_{m\bk+\bq}(T)$ the Fermi-Dirac distribution, $\eta$ a positive real infinitesimal value for causality, $N_q$ the total number of phonon momenta used to discretize the first Brillouin Zone,  $e_{\kappa\alpha\nu}(\bq)$ the mass-scaled eigendisplacement vector of atom $\kappa$ in the Cartesian direction $\alpha$ due to a phonon mode $\nu$, and  $g_{mn\kappa\alpha}(\bk,\bq)$ the electron-phonon matrix element given by
\begin{equation}\label{eq:g}
g_{mn\kappa\alpha}(\bk,\bq) = \big\langle u_{m\bk+\bq} \big| V_{\bq \kappa \alpha} | u_{n\bk} \big\rangle, 
\end{equation} 
where $ | u_{n\bk} \rangle$ is the periodic part of the ground-state electronic wavefunction with the associated nonlocal Kohn-Sham (KS) potential $V(\br,\br')$.

We note that the linearized quasiparticle spectral weight $Z_{n\bk}(T)$ from Eq.~\eqref{eq:lz} depends only on the Fan term and has a simple analytic form:
\begin{multline}
 Z^{-1}_{n\bk}(T)  = 1 + \frac{1}{N_q} \sum_{\substack{\mathbf{q}\nu m \\ \kappa\alpha\kappa'\beta}} \frac{1}{2 \omega_{\bq\nu}} g_{mn\kappa\alpha}^*(\bk,\bq) g_{mn\kappa'\beta}(\bk,\bq) \\
 \times \Bigg\{[n_{\bq\nu}(T) + 1 - f_{m\bk+\bq}(T)]\frac{(\varepsilon_{n\bk} \!-\! \varepsilon_{m\bk+\bq} \!-\! \omega_{\bq\nu})^2 \!-\! \eta^2 }{\bigl[(\varepsilon_{n\bk} \!-\! \varepsilon_{m\bk+\bq} - \omega_{\nu\bq})^2 \!+\! \eta^2 \bigr]^2 } \\ 
 \!\!\!
 + \! [n_{\bq\nu}(T) \!+\! f_{m\bk+\bq}(T)]\frac{(\varepsilon_{n\bk} \!-\! \varepsilon_{m\bk+\bq} \!+\! \omega_{\bq\nu})^2 \!-\! \eta^2 }{\bigl[(\varepsilon_{n\bk} \!-\! \varepsilon_{m\bk+\bq} \!+\! \omega_{\bq\nu})^2 \!+\! \eta^2 \bigr]^2}  \Bigg\}.
\end{multline}

The adiabatic version of the Fan self-energy corresponds to assuming that energy differences are larger than the characteristic phonon frequencies $E-\varepsilon_{m\bk} \gg \omega_{\bq\nu}$ which gives
\begin{multline}\label{eq:fanadi}
\Sigma_{nn'\bk}^{\rm Fan, ad}(E, T) = \frac{1}{N_q} \sum_{\substack{\mathbf{q}\nu m \\ \kappa\alpha\kappa'\beta}} \frac{g_{mn\kappa\alpha}^*(\bk,\bq)g_{mn'\kappa'\beta}(\bk,\bq)}{2 \omega_{\bq\nu}} \\
\times  e_{\kappa\alpha\nu}^*(\bq)e_{\kappa'\beta\nu}(\bq) \frac{2n_{\bq\nu}(T) + 1}{E-\varepsilon_{m\bk+\bq} + i\eta}.
\end{multline}

Instead, the exact DW term is static, adiabatic and real:
\begin{multline}
\Sigma_{nn'\bk}^{\rm DW}(T) = \frac{1}{N_q} \sum_{\mathbf{q}\nu} \frac{n_{\bq\nu}(T)+1/2}{2 \omega_{\bq\nu}} \mathcal{D}_{nn'\nu}(\bk,\bq),
\end{multline}
where 
\begin{multline}
\mathcal{D}_{nn'\nu}(\bk,\bq) \equiv \sum_{\kappa\alpha\kappa'\beta} \big\langle u_{n\bk} \big| \partial_{-\bq \kappa \alpha}\partial_{\bq \kappa' \beta} \hat{V}^{\rm KS} | u_{n'\bk} \big\rangle \\
\times e_{\kappa\alpha\nu}^*(\bq)e_{\kappa'\beta\nu}(\bq). 
\end{multline}
In practice, the term $\mathcal{D}$ is difficult to compute from perturbation theory as it involves a second-order derivative of the KS potential.

The simplest approximation to avoid computing the DW term would be to simply neglect the DW self-energy and keep only the Fan term:
\begin{equation} \label{eq:approx_Fan_only}
    \Sigma_{nn\bk}(E, T) \approx \Sigma^{\rm Fan}_{nn\bk}(E, T).
\end{equation}
However, this approximation is very inaccurate in practice.
In fact, the Fan and DW terms separately are large, momentum dependent, and of opposite sign~\cite{Ponce2014}.
The cancellation between the two is crucial.

A better approximation~\cite{Ponce2016a} is motivated by Luttinger's theorem~\cite{Luttinger1960b}, which states that at $T=0$, the volume enclosed by the Fermi surface must be conserved after renormalization.
For a homogeneous electron gas, this theorem implies
\begin{equation}
\Re[\Sigma_{\bk_{\rm F}}^{\rm Fan}(\varepsilon^{\rm F}, 0)] + \Sigma_{\bk^{\rm F}}^{\rm DW}(0) = 0,
\end{equation}
where $\varepsilon^{\rm F}$ is the Fermi energy and $\bk^{\rm F}$ is the Fermi wavevector.
This equation relates the $T=0$ DW self-energy at the Fermi wavevector with the Fan self-energy:
\begin{equation} \label{eq:luttinger_DW}
    \Sigma_{\bk_{\rm F}}^{\rm DW}(0) = -\Re[\Sigma_{\bk^{\rm F}}^{\rm Fan}(\varepsilon^{\rm F}, 0)], 
\end{equation}
which gives what we called the ``Luttinger approximation''~\cite{Ponce2016a, Giustino2017}
\begin{equation}\label{eq:luttinger}
    \Sigma_{nn'\bk}^{\rm Lut}(E, T) \equiv \Sigma_{nn'\bk}^{\rm Fan}(E, T) - \Sigma_{\bk^{\rm F}}^{\rm Fan}(\varepsilon^{\rm F}, 0),
\end{equation}
where in the case of insulators $\bk^{\rm F}$ is the momentum location of the valence band maximum (VBM) or the conduction band minimum (CBM), and in the case of metals it is the average Fermi wavevector. 
We note that here the Luttinger approximation uses \Eq{eq:luttinger_DW} in regimes where it is not strictly valid: solids, finite temperatures, multiple bands, and wavevectors away from the Fermi surface.

Interestingly, one can make a further approximation to Eq.~\eqref{eq:luttinger} [see Eqs.~\eqref{eq:luttingerA}-\eqref{eq:approx_FanA}] which then gives reasonable (albeit uncontrolled) results. 
That approximation can be used to study mass enhancement of semiconductors and its validity will be studied later in this paper.
%
In any case, for zero-point renormalization and change of bandgap with temperature, one cannot rely on this approximation and needs to find an alternative approximation for the DW term.

\subsection{Rigid-ion approximation}
To overcome the challenge in computing $\mathcal{D}$ from perturbation theory, the common strategy is to use the rigid-ion approximation (RIA)~\cite{Allen1976,Ponce2014a} to express the second-order matrix elements in terms of first-order one. 
This approximation has been found to be quite accurate for solids~\cite{Ponce2014a,Mann2024} but to fail in molecules~\cite{Gonze2011}. 
The diagonal part of the DW term within the RIA can be written as~\cite{Ponce2015}
\begin{multline}\label{eq:dwselria}
    \Sigma_{nn\bk}^{\rm DW, RIA}(T) = \frac{1}{N_q} \sum_{\mathbf{q}\nu} \frac{n_{\bq\nu}(T)+1/2}{2 \omega_{\bq\nu}}
    \mathcal{D}^{\rm RIA}_{nn\nu}(\bk,\bq),
\end{multline}
where
\begin{multline}\label{eq:dwria}
\mathcal{D}_{nn\nu}^{\rm RIA}(\bk,\bq) \equiv - \sum_{\substack{m\neq n \\ \kappa\alpha\kappa'\beta}} g_{mn\kappa\alpha}^*(\bk,\boldsymbol{\Gamma}) g_{mn\kappa'\beta}(\bk,\boldsymbol{\Gamma})  \\
\times \frac{e_{\kappa\alpha\nu}^*(\bq)e_{\kappa\beta\nu}(\bq) + e_{\kappa'\alpha\nu}^*(\bq)e_{\kappa'\beta\nu}(\bq)}{\varepsilon_{n\bk} - \varepsilon_{m\bk}}. 
\end{multline}
We note that in the fourth line of Eq.~(16) in Ref.~\onlinecite{Ponce2015}, the authors erroneously made the DW term in the RIA dynamical and the frequency dependence $\omega$ at the denominator of that term should be replaced by $\varepsilon_{n\bk}$, as correctly done here in Eq.~\eqref{eq:dwria}.
An alternative form, developed in Ref.~\onlinecite{Lihm2020}, gives access to the off-diagonal elements:
\begin{align}\label{eq:dwria_offdiag}
    \mathcal{D}_{nn'\nu}^{\rm RIA}&(\bk,\bq) \equiv  \nonumber \\
   & i \!\! \sum_{\kappa\alpha\beta} e_{\kappa\alpha\nu}^*(\bq)e_{\kappa\beta\nu}(\bq)  \langle u_{n\bk} | [ V_{\boldsymbol{\Gamma}\kappa\alpha}, \hat{p}_\beta ] | u_{n'\bk} \rangle   \\
    = & i \!\! \sum_{m\kappa\alpha\beta} e_{\kappa\alpha\nu}^*(\bq)e_{\kappa\beta\nu}(\bq) \big\{ \langle u_{n\bk} | V_{\boldsymbol{\Gamma}\kappa\alpha}  | u_{m\bk} \rangle \langle u_{m\bk} | \hat{p}_\beta | u_{n'\bk} \rangle \nonumber  \\
    &   - \langle u_{n\bk} | \hat{p}_\beta | u_{m\bk} \rangle \langle u_{m\bk} | V_{\boldsymbol{\Gamma}\kappa\alpha}  | u_{n'\bk} \rangle \big\}\\
    = & i \!\! \sum_{m\kappa\alpha\beta} e_{\kappa\alpha\nu}^*(\bq)e_{\kappa\beta\nu}(\bq)  \big\{ g_{nm\kappa\alpha}(\bk,\boldsymbol{\Gamma}) p_{mn'\beta}(\bk) \nonumber \\
    &  - p_{nm\beta}(\bk) g_{mn'\kappa\alpha}(\bk,\boldsymbol{\Gamma}) \big\},
\end{align}
where $p_{mn\alpha}(\bk)$ is the matrix element of the momentum operator $\hat{p}_\alpha = -i\partial / \partial r_\alpha$.

Overall, the combination of Eqs.~\eqref{eq:rs}, \eqref{eq:fanadi}, and \eqref{eq:dwselria} form the original adiabatic AHC theory.
However, it was recognized in 2015~\cite{Ponce2015,Miglio2020} that such formulation would break down in infra-red active materials and that 
a non-adiabatic version of the AHC should be used instead. 
The non-adiabatic version is obtained by combining Eqs.~\eqref{eq:rs}, \eqref{eq:fan}, and \eqref{eq:dwselria}. 

\subsection{Diamond and BAs}

To illustrate various technical points made in this work and quantify the level of agreement between first-principles codes, we chose two representative simple materials: diamond and BAs.
Diamond was chosen as one of the few infrared (IR)-inactive materials and because it is one of the most studied materials with a large zero-point renormalization of its direct and indirect bandgaps~\cite{Logothetidis1992}.
BAs was chosen as a representative example of IR-active materials due to its exceptional thermal ($\sim$1000~W/mK) and carrier ($\sim$1500~cm$^2$/Vs) transport~\cite{Liu2018,Yue2022,Shin2022,Esho2023} as well a negligible impact of thermal expansion on the band renormalization due to covalent bonding~\cite{Bravi2019}.
In addition, zero-point renormalization and temperature dependence of the electronic bandstructure have not yet been reported experimentally for BAs and we found only one prior theoretical report using a frozen-phonon method~\cite{Bravi2019}.

Care was taken to perform exactly the same calculations in each code with the same computational parameters.
In particular, thanks to the recent work by Dr.\ Matteo Giantomassi, the \textsc{Abinit} code can now read pseudopotential in the unified pseudopotential format (UPF) such that the same pseudopotential was used in each code. 
We used modified versions of the norm-conserving pseudopotential from the \textsc{PseudoDojo}~\cite{Setten2017} library v0.4.1 with standard accuracy and the PBE exchange-correlation functional where the non-linear core correction (NLCC) was removed since the linear response implementation of dynamical quadrupole in the \textsc{Abinit} software does not support it yet~\cite{Royo2019}. 
The following valence electrons 2s$^2$2p$^2$ for carbon, 2s$^2$2p$^1$ for boron, and 3d$^{10}$4s$^2$4p$^3$ for arsenic, were explicitly treated.

To focus on the electron-phonon renormalization, we neglect thermal expansion and perform all calculations at a fixed lattice parameter of 6.7035~Bohr for diamond and 9.0850~Bohr for BAs.
DFT yields an indirect and direct bandgap for diamond of 4.2~eV and 5.7~eV, respectively, and for BAs of 1.2~eV and 3.3~eV, respectively.
The comparison between \textsc{Abinit} and \textsc{Quantum ESPRESSO} for the total energy, bandgaps, dielectric constant, Born effective charge, and zone center phonon frequencies are given in Table S1 of the SI~\cite{SI}.
The quadrupole tensor is computed using linear response with \textsc{Abinit}~\cite{Stengel2013} and used both in \textsc{Abinit} and \textsc{Quantum ESPRESSO}.
In addition, we compare in Fig.~S1 of the SI~\cite{SI} the electron and phonon bandstructure along high-symmetry lines between \textsc{Abinit}, \textsc{Quantum ESPRESSO} and \textsc{EPW}. 
Overall, the agreement is good with the largest difference (4\%) being observed for the dielectric constant $\epsilon^{\infty}$ between the codes.

\subsection{Empty state convergence}\label{sec:sos}

\begin{figure}[t]
  \centering
  \includegraphics[width=0.9\linewidth]{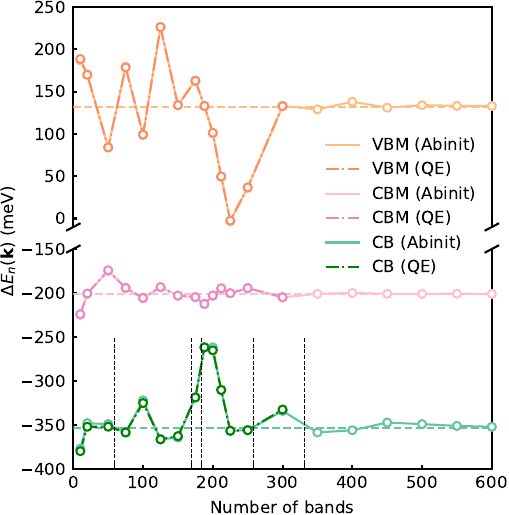}
  \caption{\label{fig:sos}
  \textbf{Band convergence of the zero-point renormalization in diamond.}
Convergence with respect to the number of bands in the sum-over-state approach of the zero-point renormalization (ZPR) in diamond at the valence band maximum (VBM), conduction band at \textbf{k}=$\Gamma$ (CB), and conduction band minimum (CBM), see Fig.~S1 of the SI~\cite{SI}.
We also show the ZPR computed with the Sternheimer method using 9 bands in the active space (horizontal dashed lines) obtained using \textsc{Abinit}.
The largest relative difference between \textsc{Abinit} and \textsc{Quantum ESPRESSO} across all data points is 1.1~\%.
Dashed vertical lines indicate the place (upper band index) where there is an energy gap in the electronic bandstructure at the zone center, see Fig.~\ref{fig:bandsAb}.
}
\end{figure}

In practice, when solving Eqs.~\eqref{eq:fan} and ~\eqref{eq:dwria}, the convergence with respect to the number of bands $m$ is extremely slow. 
For example, as seen in Fig.~\ref{fig:sos} for the case of diamond, one needs 600 bands to converge the ZPR of the bandgaps. 
To our knowledge, the slow convergence of the sum-over-state approach was never reported in a journal paper. 
However we note that P. Boulanger found in his PhD thesis that the second-order derivative of the VBM eigenvalues of silicon required over 500 bands to be converged with the sum-over-state approach~\cite{Boulanger2010}.

To overcome this challenge, the sum-over-state expressions in Eqs.~\eqref{eq:fan} and ~\eqref{eq:dwria} resulting from the first-order perturbed wavefunction can be replaced by the iterative solutions to linear Euler-Lagrange equations. 
These equations are often called Sternheimer equations and are commonly used in DFPT~\cite{Gonze1997,Baroni2001,Gonze2011}.
However, these linear equations formally correspond to the adiabatic Fan self-energy, Eq.~\eqref{eq:fanadi} and not the non-adiabatic version Eq.~\eqref{eq:fan}.

This is a major issue as adiabatic formulation diverges in IR-active materials~\cite{Ponce2015,Miglio2020}. 
To circumvent it, the self-energies are divided into a non-adiabatic sum-over-state expression on a small ``active'' space and an adiabatic Sternheimer formulation 
corresponding to the contribution of the infinite number of empty states above the active space A which yields the complete renormalization using the RIA:
\begin{widetext}
\begin{subequations} \label{eq:sterneq}
\begin{align}
    \label{eq:sterneq_sum}
    \Delta E_{n\bk}(T)
    &= \frac{1}{N_q} \sum_{\bq \nu}
    \Bigl[ \Delta E^{\rm Fan,\,A}_{n\bk; \bq\nu}(T)
    + \Delta E^{\rm Fan,\,R}_{n\bk, \bq\nu}(T)
    + \Delta E^{\rm DW,\,A}_{n\bk, \bq\nu}(T)
    + \Delta E^{\rm DW,\,R}_{n\bk, \bq\nu}(T) \Bigr],
    \\
    \label{eq:sterneq_FanA}
    \Delta E^{\rm Fan,\,A}_{n\bk, \bq \nu}(T)
    &= \Re \frac{1}{2 \omega_{\bq\nu}} \sum_{\kappa\alpha\kappa'\beta}
    \sum_{m}^{\rm A} g_{mn\kappa\alpha}^*(\bk,\bq) g_{mn\kappa'\beta}(\bk,\bq)  \sum_{\pm} \frac{n_{\bq\nu}(T) + \tfrac{1}{2} \pm \bigl[ f_{m\bk+\bq}(T) - \tfrac{1}{2} \bigr]}{\varepsilon_{n\bk}-\varepsilon_{m\bk+\bq} \pm \omega_{\bq\nu} + i\eta}
    e_{\kappa\alpha\nu}^*(\bq)e_{\kappa'\beta\nu}(\bq) ,
    \\
    \label{eq:sterneq_FanR}
    \Delta E^{\rm Fan,\,R}_{n\bk, \bq \nu}(T)
    &= 2\Re \frac{1}{2 \omega_{\bq\nu}} \sum_{\kappa\alpha\kappa'\beta}
    \langle P_{\bk+\bq}^{\rm c} \Delta_{\bq \kappa \alpha} u_{n\bk} | V_{\bq \kappa'\beta} | u_{n\bk} \rangle
    e_{\kappa\alpha\nu}^*(\bq)e_{\kappa'\beta\nu}(\bq)
    \bigl[ n_{\bq\nu}(T) + \tfrac{1}{2} \bigr] ,
    \\
    \label{eq:sterneq_DWA}
    \Delta E^{\rm DW,\,A}_{n\bk, \bq \nu}(T)
    &= -\Re \frac{1}{2 \omega_{\bq\nu}} \sum_{\kappa\alpha\kappa'\beta}
    \sum_{m\neq n}^{\rm A} \frac{g_{mn\kappa\alpha}^*(\bk,\boldsymbol{\Gamma}) g_{mn\kappa'\beta}(\bk,\boldsymbol{\Gamma})}{\varepsilon_{n\bk} - \varepsilon_{m\bk}}
     \big( e_{\kappa\alpha\nu}^*(\bq)e_{\kappa\beta\nu}(\bq) \!+\! e_{\kappa'\alpha\nu}^*(\bq)e_{\kappa'\beta\nu}(\bq) \big) \bigl[ n_{\bq\nu}(T) \!+\! \tfrac{1}{2} \bigr],
    \\
    \label{eq:sterneq_DWR}
    \Delta E^{\rm DW,\,R}_{n\bk, \bq \nu}(T)
    &= -\Re \frac{1}{2 \omega_{\bq\nu}} \sum_{\kappa\alpha\kappa'\beta}
    \langle P_{\bk}^{\rm c} \Delta_{\boldsymbol{\Gamma} \kappa \alpha} u_{n\bk} | V_{\boldsymbol{\Gamma} \kappa' \beta} | u_{n\bk} \rangle
     \big( e_{\kappa\alpha\nu}^*(\bq)e_{\kappa\beta\nu}(\bq) + e_{\kappa'\alpha\nu}^*(\bq)e_{\kappa'\beta\nu}(\bq) \big) \bigl[ n_{\bq\nu}(T) + \tfrac{1}{2} \bigr].
\end{align}
\end{subequations}
\end{widetext}
Here, $P_{\bk+\bq}^{\rm c}$ is the complimentary projector to the active subspace A,
and $| P_{\bk+\bq}^{\rm c} \Delta_{\bq \kappa\alpha} u_{n\bk} \rangle$ is the solution to the Sternheimer equation~\cite{Gonze2011}
\begin{equation}
    (H_{\bk+\bq} - \varepsilon_\nk) | P_{\bk+\bq}^{\rm c} \Delta_{\bq \kappa\alpha} u_{n\bk} \rangle
    = - P_{\bk+\bq}^{\rm c} V_{\bq \kappa\alpha} | u_\nk \rangle\,.
\end{equation}

Using translational invariance, the active and rest DW terms within the RIA can be exactly rewritten as~\cite{Lihm2020}
\begin{align}
    \Delta E^{\rm DW,\,A}_{n\bk, \bq \nu}(T) =&  \Re \frac{1}{2 \omega_{\bq\nu}} \sum_{\kappa\kappa'\alpha\beta}
    \sum_{m}^{\rm A} i g_{mn\kappa\alpha}^*(\bk,\boldsymbol{\Gamma}) p_{mn\beta}(\bk) \delta_{\kappa\kappa'}  \nonumber \\
  \!  \times \big( e_{\kappa\alpha\nu}^*(\bq)e_{\kappa\beta\nu}(\bq)  + & e_{\kappa'\alpha\nu}^*(\bq)e_{\kappa'\beta\nu}(\bq) \big)  \bigl[ n_{\bq\nu}(T) \!+\! \tfrac{1}{2} \bigr] \! \! \label{eq:sterneq_DWA_p} \\
    \Delta E^{\rm DW,\,R}_{n\bk, \bq \nu}(T)
    =&  \Re \frac{1}{2 \omega_{\bq\nu}} \sum_{\kappa\kappa'\alpha\beta}
    i \langle u_{n\bk} | V_{\boldsymbol{\Gamma} \kappa \alpha} P_{\bk}^{\rm c} p_{\beta} | u_{n\bk} \rangle \delta_{\kappa\kappa'}  \nonumber\\
   \!\! \times\big( e_{\kappa\alpha\nu}^*(\bq)e_{\kappa\beta\nu}(\bq)  + & e_{\kappa'\alpha\nu}^*(\bq)e_{\kappa'\beta\nu}(\bq) \big)  \bigl[ n_{\bq\nu}(T) \!+\! \tfrac{1}{2} \bigr], \! \label{eq:sterneq_DWR_p}
\end{align}
where the matrix element $\langle \psi_{n\bk} | V_{\boldsymbol{\Gamma} \kappa \alpha} P_{\bk}^{\rm c} p_{\beta} | \psi_{n\bk} \rangle$ in Eq.~\eqref{eq:sterneq_DWR_p} can be easily calculated in the plane-wave basis and where, due to the RIA, the two phonon eigendisplacement terms in Eqs.~\eqref{eq:sterneq_DWA_p} and \eqref{eq:sterneq_DWR_p} can be written with a factor two and the same atomic index $\kappa$.

In the case of a small active space region, and especially relevant in the context of Wannier function downfolding, it has been suggested~\cite{Ponce2016a} that 
the Luttinger approximations of Eq.~\eqref{eq:luttinger} could be performed in the active subspace only:   
\begin{equation}\label{eq:luttingerA}
\Sigma_{n\bk}^{\rm Lut, A}(E, T) \equiv \Sigma_{n\bk}^{\rm Fan, A}(E, T) - \Re[\Sigma_{\bk^{\rm F}}^{\rm Fan}(\varepsilon^{\rm F}, T)],
\end{equation}
which could perform better for intrinsic semiconductors due to error cancellation but does depend on the size of the active space chosen.

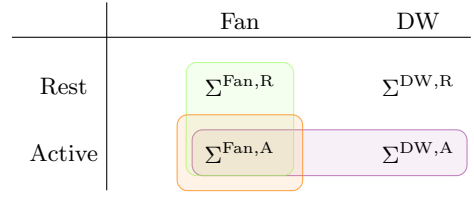
\begin{figure}[tb]
\begin{tikzpicture}[]
\matrix[row sep=0.4cm, column sep=1.2cm]{
    \node(x0){}      ; & \node{Fan}                      ; & \node(x2) {DW}; \\
    \node(y1){Rest}  ; & \node(a) {$\Sigma^{\rm Fan,R}$} ; & \node(c) {$\Sigma^{\rm DW,R}$}; \\
    \node(y2){Active}; & \node(b) {$\Sigma^{\rm Fan,A}$} ; & \node(d) {$\Sigma^{\rm DW,A}$}; \\
};
\draw [black] ([xshift=-5.4cm] x2.south) -- ([xshift= 0.7cm] x2.south);
\draw [black] ([yshift= 2.0cm] y2.east)  -- ([yshift=-0.5cm] y2.east);
\begin{scope}[on background layer,rounded corners]
\node [fit=(b) (d),purpleish, inner xsep= 0.5mm, inner ysep= 0.5mm ]{};
\node [fit=(a) (b),greenish,  inner xsep= 1.3mm, inner ysep= 0.5mm ]{};
\node [fit=(b) (b),orangeish, inner xsep= 2.5mm, inner ysep= 2.5mm]{};
\end{scope}
\end{tikzpicture}
\caption{\textbf{Decomposition of the Allen-Heine-Cardona self-energy}. It is split into an active (A) and rest (R) space.  
The decomposition allows to define four practical approximations to the total self-energy $\Sigma_{n\bk}$: (i) the Fan only term $\Sigma_{n\bk}^{\rm Fan}$ with Eq.~\eqref{eq:approx_Fan_only} (green box), (ii) active space only $\Sigma_{n\bk}^{\rm A}$ using Eq.~\eqref{eq:approx_A} (mauve box), (iii) the Fan term on the active space only $\Sigma_{n\bk}^{\rm Fan,A}$ using Eq.~\eqref{eq:approx_FanA} (orange box), as well as (iv) the Luttinger approximation $\Sigma_{n\bk}^{\rm Lut, A}$ of Eq.~\eqref{eq:luttingerA}.  
}
\label{fig:decomp}
\end{figure}

Based on this decomposition, one can make further approximations to the self-energy.
The idea is that the rest-space contribution to the Fan and DW self-energies are both large and approximately cancel each other.
Then, one may approximate by dropping the rest-space contribution~\cite{Engel2020}:
\begin{equation} \label{eq:approx_A}
    \Sigma_{n\bk}^{\rm A}(E, T) \equiv \Sigma_{n\bk}^{\rm Fan,A}(E, T) + \Sigma_{n\bk}^{\rm DW,A}(E, T).
\end{equation}
One may further neglect the active-space DW term to find
\begin{equation} \label{eq:approx_FanA}
    \Sigma_{n\bk}(E, T) \approx \Sigma_{n\bk}^{\rm Fan,A}(E, T).
\end{equation}
\Figu{fig:decomp} summarizes the five decompositions of the self-energy and the approximations we consider.
One of the goals of this work is to assess the accuracy of these five approximations when computing the electron-phonon mass enhancement.

Coming back to Eq.~\eqref{eq:sterneq}, in the case of diamond, we choose an active space composed of 9 bands, and the rest space contribution is computed with the Sternheimer equation.
We show the results in Fig.~\ref{fig:sos} with horizontal dashed lines, demonstrating a perfect agreement with the converged sum-over-state approach, with a significant computational advantage. 
We note that we have here used the interpolation of the perturbed potential~\cite{Gonze2020} to speedup the $\mathbf{q}$-point convergence.
Such validation of the sum-over-state approach with respect to the Sternheimer approach has never been reported in that case so far.

\begin{figure}[t]
  \centering
  \includegraphics[width=0.99\linewidth]{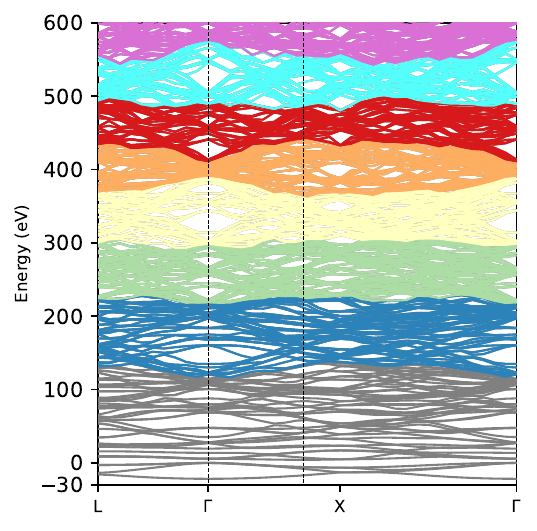} 
  \caption{\label{fig:bandsAb}
\textbf{Electronic bandstructure of diamond}. Each color (gray, blue, green, yellow, orange, red) indicates a block of 50 bands and where the vertical dashed lines show the position of the valence band maximum (VBM) and conduction band minimum, respectively.    
The VBM is located at 0~eV on the energy axis. 
}
\end{figure}

Remarkably, the CBM which is located at 72\% of the $\Gamma$-$X$ high-symmetry direction, converges much faster than the VBM and conduction band (CB) at $\Gamma$, requiring only 100 bands. 
The reason for this fast convergence can be found by analyzing the electronic bandstructure of diamond, shown in Fig.~\ref{fig:bandsAb}.
At $\Gamma$, large energy gaps can be seen as a function of energy. 
Due to the importance of the $\bq=\boldsymbol{\Gamma}$ momentum in the DW term, these gaps result in pinning of the bands above and below, which creates over- and under-estimations of the ZPR when solving the sum-over-state equation. 
In contrast, these energy gaps are absent at the CBM location which explains the faster convergence.  
In both Figs.~\ref{fig:sos} and \ref{fig:bandsAb} we use a larger plane-wave truncation of 60~Ha to be able to describe all the bands. 
For this comparison, we use a non-converged Fourier-interpolated 40$\times$40$\times$40 \textbf{q}-point grid with a CBM located at 0.8X instead of the true CBM located at 0.72X in diamond. 
We explain in more detail the Fourier interpolation of the perturbed potential laterwards.

In the case of BAs, presented in Figs.~S2 and S3 of the SI~\cite{SI}, the sum-over-state convergence is even slower and reaches the Sternhimer values only after 1200 bands have been included. 
The reason for this overall slower convergence is due to the fact that the density of bands per energy is higher than in the case of diamond, which explains why more bands are needed to converge the sum-over-state expressions.
Also in this case, the CBM which is located at 80\% the $\Gamma$-$X$ high-symmetry direction, converges faster than the states at the zone center for the same reason as diamond.

\subsection{Momentum integral convergence}
\label{sec:mom}

The convergence of the $\bq$-integral with respect to grid density has been shown to be slow~\cite{Ponce2014}.
There exist at least three strategies to converge the $\bq$ integration: (i) direct explicit calculation of response function at finite $\bq$, using the irreducible Brillouin zone; (ii) interpolation of the perturbed potential; or (iii) interpolation of the electron-phonon matrix elements using WFPT.

\begin{figure}[t]
  \centering
  \includegraphics[width=0.85\linewidth]{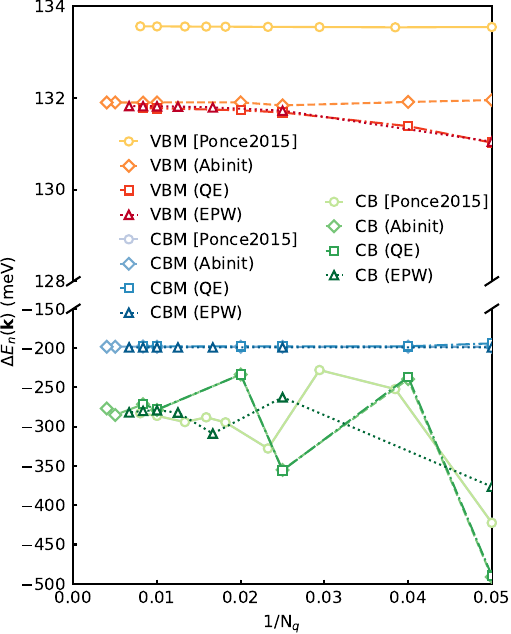} 
  \caption{\label{fig:qconv}
\textbf{Momentum convergence of diamond.}  
Convergence of the valence band maximum (VBM), which is at $\Gamma$, conduction band minimum (CBM) at 0.72$\Gamma X$ and conduction band at the zone center (CB) of the non-adiabatic Rayleigh-Schr\"odinger zero-point motion renormalization (0~K) as a function of inverse linear $\bq$ momentum density such that the grid size is $N_q \times N_q \times N_q$ and $1/N_q = 0$ corresponds to an infinite grid density.  
We compare the interpolation of the perturbed potential including quadrupoles, 9 active bands coupled with Sternheimer, an 8$\times$8$\times$8 coarse $\bq$ grid, and a 5~meV fixed smearing as implemented in the \textsc{Abinit} and \textsc{Quantum ESPRESSO} software 
as well as interpolation of the Wannier interpolation using Wannier function perturbation theory as implemented in the \textsc{EPW} code with 8 bands, an 8$\times$8$\times$8 coarse $\bq$ grid, and a 5~meV fixed smearing.
The data with circles are from Ref.~\onlinecite{Ponce2015}, made 10 years ago, using direct calculation with \textsc{Abinit} v7.11 with a different pseudopotential (Perdew and Zunger parametrization of LDA), a 10~meV smearing, a different lattice parameter of 6.652~Bohr and nonetheless show a remarkable agreement.
Quadrupoles are intrinsically included since there are no interpolation methods used. 
The convergence with the adiabatic frozen-phonon approach is presented separately in Fig.~\ref{fig:qconvZG} due to quite different axis scales.
}
\end{figure}

In the first case, the convergence is computationally demanding but possible since we are looking at the renormalized bandstructure for a few selected $\bk$ points. 
In addition, it was shown~\cite{Ponce2015} that the non-adiabatic ZPR of IR-inactive materials has a typical rapid polynomial convergence while the IR-active materials converge linearly on an inverse momentum-density scale. 
That systematic convergence behavior allows for an efficient extrapolation of the results to infinite $\bq$ density.  
This strategy was used in the following works~\cite{Ponce2014,Ponce2014a,Ponce2015,Antonius2015,Miglio2020,BrousseauCouture2020,BrousseauCouture2023} and, in particular, we show the 
results of Ref.~\onlinecite{Ponce2015} in Fig.~\ref{fig:qconv} for the case of the non-adiabatic solution for diamond.
The results converge smoothly with direct integration for the band edges but oscillate more strongly for the conduction band (CB) at the zone center due to energy resonances with other parts of the Brillouin zone.

In the second approach, we interpolate the perturbed potential $V_{\bq\kappa\alpha}$ from Eq.~\eqref{eq:g}. 
This approach is based on the seminal idea of Eiguren and Ambrosch-Draxl, developed as a private Fortran code~\cite{Eiguren2008}, then in \textsc{Abinit}~\cite{Gonze2016,Gonze2020,BrousseauCouture2022} and recently also implemented in \textsc{Quantum ESPRESSO}~\cite{Giannozzi2017,Lihm2020}.
For the interpolation to be accurate, the Fourier transform of the perturbed potential needs to be short-ranged. 
For this reason, one first needs to separate the perturbed potential into nonlocal (NL), short-ranged ($\mathcal{S}$) and long-ranged ($\mathcal{L}$) parts:
\begin{multline}\label{eq:separations}
V_{\bq\kappa\alpha}(\br,\br') = 
V_{\bq\kappa\alpha}^{\rm NL}(\br,\br') \\
+ \delta(\br - \br') \bigl[
V_{\bq\kappa\alpha}^{\mathcal{S}}(\br) + V_{\bq\kappa\alpha}^{\mathcal{L}}(\br) \bigr].
\end{multline} 
The expansion of the long-range perturbed potential to order $\mathcal{O}(\abs{\bq}^0)$ gives~\cite{Brunin2020a,Brunin2020,Ponce2023,Ponce2023a}:
\begin{multline}\label{eq:vlr_lw}
V_{\mathbf{q}\kappa\alpha}^{\mathcal{L}}(\mathbf{r}) = \frac{4\pi e}{\Omega} \frac{f(|\mathbf{q}|)}{|\mathbf{q}|^2\tilde{\epsilon}(\mathbf{q})} e^{-i \mathbf{q} \cdot \boldsymbol{\tau}_\kappa} \bigg[    \sum_\beta i q_\beta Z_{\kappa\alpha\beta}   \\
+ \sum_\gamma \frac{q_\beta  q_\gamma}{2}  Q_{\kappa\alpha\beta\gamma}  \bigg] e^{i\bq \cdot \br}(1+iq_\alpha V^{\textrm{Hxc},\boldsymbol{\mathcal{E}}}(\mathbf{r}) ) ,
\end{multline}
where $\mathbf{Z}$ are the Born effective charge, $\mathbf{Q}$ dynamical quadrupoles, $V^{\textrm{Hxc},\boldsymbol{\mathcal{E}}}(\mathbf{r})$ the self-consistent potential induced by a uniform electric field $\mathcal{E}$ which is here neglected, $\Omega$ the unit-cell volume, and where the effective macroscopic dielectric function is:
\begin{equation}\label{eq:epsil}
\tilde\epsilon(\mathbf{q}) =  \frac{ \mathbf{q}\cdot \boldsymbol{\varepsilon}\cdot\mathbf{q}}{\phantom{^2}|\mathbf{q}|^2}  f(|\mathbf{q}|) +  1 -  f(|\mathbf{q}|)~.
\end{equation}
where $f(|\mathbf{q}|) = e^{-\frac{|\bq|^2 L^2}{4}}$ the range separation function and $L$ the range separation parameter that we here take $L=\frac{a}{2\pi}$ where $a$ is the lattice parameter.
Equations~\eqref{eq:vlr_lw} and \eqref{eq:epsil} as well as $f$ can be extended for 2D materials~\cite{Royo2021,Ponce2023}.

The short-ranged potential is Fourier transformed from a coarse $\bq$-point grid to real space:
\begin{equation}
W_{\kappa\alpha}(\br,\bR_{p'}) = \frac{1}{N_{p'}} \sum_{\bq} e^{-i\bq \cdot(\bR_{p'} - \br)} V_{\bq\kappa\alpha}^{\mathcal{S}}(\br,\br'),
\end{equation}
where, following Ref.~\onlinecite{Ponce2021}, we use $p'$ to denote vibrational supercells and $p$ (used later) to denote electronic supercells. $N_{p'}$ is the number of unit cells in the corresponding Born-von K\'arman supercell.
One can then perform a subsequent Fourier interpolation of the short-ranged real-space potential to an arbitrary $\bq'$-point
\begin{equation}
V_{\mathbf{q}'\kappa\alpha}^{\mathcal{S}}(\mathbf{r}) = \sum_{\bR_{p'}} e^{i\bq' \cdot (\bR_{p'} - \br)} W_{\kappa\alpha}(\br,\bR_{p'}),
\end{equation}
and add back the nonlocal and long-range potentials:
\begin{multline}\label{eq:longpot}
V_{\bq'\kappa\alpha}(\br, \br') = \\
V_{\bq'\kappa\alpha}^{\rm NL}(\br, \br')
+ \delta(\br - \br') \bigl[ V_{\bq'\kappa\alpha}^{\mathcal{S}}(\br) + V_{\bq'\kappa\alpha}^{\mathcal{L}}(\br) \bigr]. 
\end{multline}

\begin{figure}[t]
  \centering
  \includegraphics[width=0.99\linewidth]{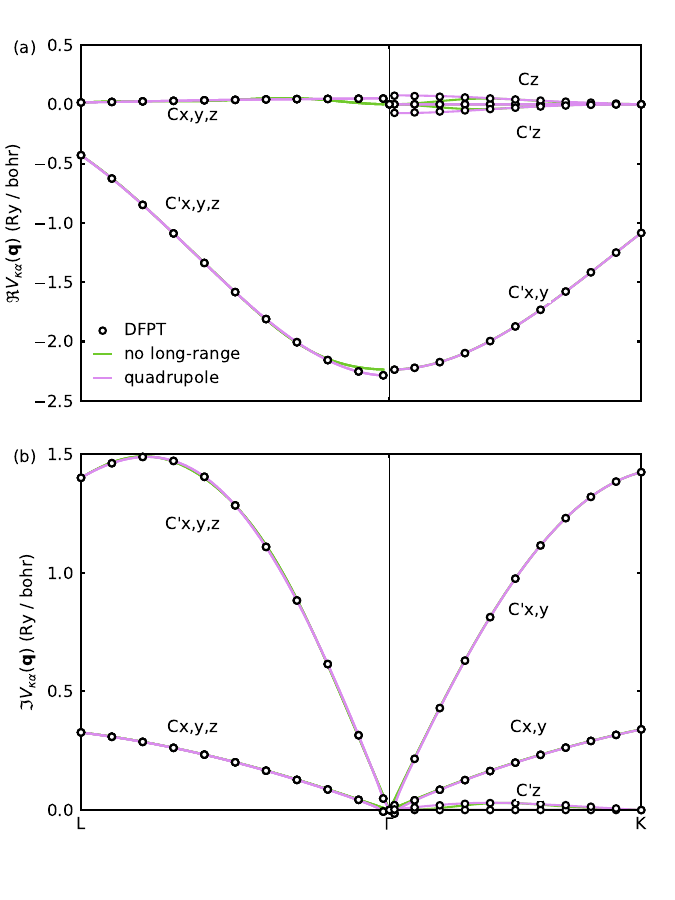} 
  \caption{\label{fig:quadqe}
\textbf{Fourier interpolated perturbed potential of diamond.}   
(a) Real and (b) imaginary part of the Fourier interpolated perturbed potential including long-range quadrupoles (mauve) $Q_{\kappa\alpha\beta\gamma}=2.46|\varepsilon_{\alpha\beta\gamma}|$~eBohr, with $\varepsilon_{\alpha\beta\gamma}$ being the Levi-Civita symbol,  or without any long-range treatments (green) compared with direct results obtained from density functional perturbation theory (empty circles).
The different branches correspond to the movement of one of two carbon atoms (C and C$^\prime$) in the indicated Cartesian direction.  
The calculation is done with \textsc{Quantum ESPRESSO} with an 8$\times$8$\times$8 coarse $\bk$ and $\bq$ grids.  
}
\end{figure}

For this work, we extended the original \textsc{Quantum ESPRESSO} implementation~\cite{Lihm2020} to support quadrupoles and show the accuracy of the interpolation in Fig.~\ref{fig:quadqe} for the case of diamond where the quadrupole tensor was computed in Ref.~\onlinecite{Ponce2021}. 
Small deviations close to the $\Gamma$-point can be observed in the figure, in particular, the two carbon atoms moving in the $z$ Cartesian direction are incorrectly degenerate when going from $\Gamma$ to $K$ if the quadrupoles are not included. 
We remark that since diamond is IR-inactive, the Born effective charges are null and there is no dipole contribution. 
In the case of BAs, dipoles play an important role in describing the behavior of the imaginary part of the interpolated perturbed potential close to the zone boundary, while the quadrupoles are crucial for a correct description of the real part: see Fig.~S4 of the SI~\cite{SI} for more information.

We compare in Fig.~\ref{fig:qconv} the convergence rate of the ZPR of the VBM, CBM and CB of diamond with respect to the $\mathbf{q}$-integration mesh using the interpolation of the perturbed potential both in \textsc{Abinit} (empty diamonds) and  \textsc{Quantum ESPRESSO} (empty squares).
We find that the agreement between codes is excellent for the band edges but is worse for other states due to numerical resonances, making it more challenging to perfectly converge the codes. 
In particular, we find that the largest relative difference at convergence is 0.26\% for the VBM and CBM, while the agreement worsens to 2.05\% for the CB but is still reasonable.
We present in Fig.~S5 of the SI~\cite{SI} the relative error as a function of integration grid density with respect to the \textsc{Abinit} converged result.  
Interestingly, for the band edges, not only are the converged results very close but close agreement is also found for each $\bq$ integration grid where we find the largest relative difference to be 2.16\% for the smallest grid of $20^3$ $\bq$-points.

In the last case, the electron-phonon matrix elements are interpolated directly in the Wannier gauge using WFPT~\cite{Lihm2021}, as implemented in the \textsc{EPW}~\cite{Ponce2016a,Lee2023} code. 
The Wannier functions (WFs) of a crystal can be written as~\cite{Marzari2012}
\begin{equation}
\ket{w_{i\mathbf{R}}} = \frac{1}{\sqrt{N_k}}\sum_{m\mathbf{k}} \ket{\psi_{m\mathbf{k}}} U_{mi\mathbf{k}} e^{-i\mathbf{k}\cdot \mathbf{R}},  
\end{equation}
where the gauge transformation matrix  $U_{mi\bk}$ rotates from Bloch wavefunction $\psi_{m\bk}$ to a WF $w_{i\bR}$ with index $i = 1$ to $N^{\rm W}$ and unit cell position $\bR$ such that the resulting WFs are spatially localized.

The corresponding Wannier function perturbations (WFPs) due to a periodic displacement of atom $\kappa$ along direction $\alpha$ with momentum $\bq$ can then be given as:
\begin{multline}\label{eq:pertWF}
\ket{ \Delta_{\qka} w_{i\bR}} =  \\
\frac{1}{\sqrt{N_k}}\sum_{m\mathbf{k}} (\hat{\mathds{1}} - \hat{P}^{\rm W}) \ket{ \Delta_\qka \psi_{m\mathbf{k}}} U_{mi\mathbf{k}} e^{-i\mathbf{k}\cdot \mathbf{R}},
\end{multline}
where $\hat{P}^{\rm W}$ is the projection operator to the Wannier subspace:
\begin{equation}
 \hat{P}^{\rm W} = \sum_{\bR} \sum_{i=1}^{N^{\rm W}} \ket{w_{i\bR}} \bra{w_{i\bR}}.
\end{equation}
For disentangled WFs~\cite{Souza2001}, the states not selected from disentanglement give an additional correction~\cite{Lihm2021}.

WFPs form a spatially localized representation of the wavefunction perturbation~\cite{Lihm2021}.
Thus, WFPs can be used to interpolate the wavefunction perturbation from a coarse grid of electron momentum to a dense grid.
The interpolated wavefunction reads
\begin{align} \label{eq:wfpt_psi_itp}
    \ket{ \Delta_\qka \psi_{n\mathbf{k}} } \! =& \! \sum_{i\bR} \Bigl[ \ket{\Delta_\qka w_{i\bR}} \! U_{in\bk}
    \!+\! \ket{w_{i\bR}} \! (\Delta_\qka U)_{in\bk} \Bigr] e^{i\mathbf{k}\cdot \mathbf{R}}
    \nonumber \\
    =& \sum_{i\bR} \ket{\Delta_\qka w_{i\bR}} U_{in\bk} e^{i\mathbf{k}\cdot \mathbf{R}} \nonumber \\
    &+ \sum_{m=1}^{N^{\rm W}} \ket{\psi_{m\mathbf{k+q}}} \frac{g_{mn\kappa\alpha}(\bk,\bq)}{\varepsilon_{n\mathbf{k}} - \varepsilon_{m\mathbf{k+q}}}.
\end{align}
We substitute \Eq{eq:wfpt_psi_itp} into \Eq{eq:sterneq_FanR} to compute the rest-space Fan term.
The matrix element to be interpolated is
\begin{multline}
    \langle P^{\rm c}_{\bk+\bq}  \Delta_\qka \psi_{n\mathbf{k}} | V_\qkpb \psi_{n\bk} \rangle  \\
    = \sum_{ij\bR_e} U_{ni\bk}^\dagger  \bra{\Delta_\qka w_{i\mathbf{0}}} V_\qkpb \ket{w_{j\bR_e}} U_{jn\bk} e^{i\bk \cdot \bR_e} \\
    + \!\! \sum_{\substack{m=1 \\ m\bk+\bq \in P^{\rm c}_{\bk+\bq}}}^{N^{\rm W}}
    \frac{g_{mn\kappa\alpha}^*(\bk,\bq) g_{mn\kappa'\alpha'}(\bk,\bq) }{\varepsilon_{n\bk} - \varepsilon_{m\bk+\bq}}\,,
\end{multline}
We note that the first term is a correction coming from the change of the Wannier function.
This term is relevant only for the real part of the self-energy and therefore can be neglected in the calculation of quasiparticle lifetime.
The matrix elements that appear in the DW term [\Eqs{eq:sterneq_DWA_p} and \eqref{eq:sterneq_DWR_p}] can be directly Wannier interpolated.

In contrast to the second method where the potential is interpolated [\Eq{eq:longpot}], here the full matrix element $g(\bk,\bq)$ is interpolated which has the benefit of allowing an efficient interpolation of both $\bk$ and $\bq$.
For the same reasons as before, the electron-phonon matrix element has to be separated into long and short-range parts for accurate interpolation. 
There are however two subtleties when interpolating the matrix element instead of the potential~\cite{Ponce2023}.
The first one is that the nonlocal part of the potential, Eq.~\eqref{eq:separations}, is included in the short-range part and Fourier interpolated. 
The second one is that the long-range matrix element reads
\begin{equation}\label{eq:glong}
g^\mathcal{L}_{mn\kappa\alpha}(\mathbf{k,q}) =
\sum_{ij} U_{mi\mathbf{k+q}} \langle u_{i\mathbf{k+q}}^{\rm W}| V_{\mathbf{q}\kappa\alpha}^{\mathcal{L}} | u_{j\mathbf{k}}^{\rm W} \rangle U_{jn\mathbf{k}}^{\dagger},
\end{equation}
where  $\ket{u_{n\mathbf{k}}} = e^{-i\mathbf{k}\cdot\mathbf{r}} \ket{\psi_{n\mathbf{k}}} =  \sum_i U_{ni\mathbf{k}}^{*} \ket{u^{\rm W}_{i\mathbf{k}}}$ is the periodic part of the Bloch eigenstate and where the matrix elements are given, using Eq.~\eqref{eq:vlr_lw}, by
\begin{multline}
    \langle u_{i\mathbf{k+q}}^{\rm W}| V_{\mathbf{q}\kappa\alpha}^{\mathcal{L}} | u_{j\mathbf{k}}^{\rm W} \rangle = 
    \frac{4\pi e}{\Omega} \frac{f(|\mathbf{q}|)}{|\mathbf{q}|^2\tilde{\epsilon}(\mathbf{q})} e^{-i \mathbf{q} \cdot \boldsymbol{\tau}_\kappa} \bigg[    \sum_\beta i q_\beta Z_{\kappa\alpha\beta}
    \\
    + \sum_\gamma \frac{q_\beta  q_\gamma}{2}  Q_{\kappa\alpha\beta\gamma}  \bigg] \langle u_{i\mathbf{k+q}}^{\rm W} | \varphi_{\mathbf{q}}^{\mathcal{S}}(\mathbf{r})| u_{j\mathbf{k}}^{\rm W} \rangle, 
\end{multline} 
where $\varphi_{\mathbf{q}}^{\mathcal{S}}(\mathbf{r})$ is the dressed short range basis function~\cite{Ponce2023}. 
One can then Taylor expand to first order the periodic part of the wavefunction in $\bq$ as
\begin{equation}\label{eq:uexp}
\langle u_{i\mathbf{k+q}}^{\rm W}| = \langle u_{i\mathbf{k}}^{\rm W}| + \sum_\alpha q_\alpha \left\langle \frac{\partial u_{i\mathbf{k}}^{\rm W}}{\partial k_\alpha}\right| + \cdots\,,
\end{equation}
to obtain
\begin{multline}\label{eq:long-range-new}
 \langle u_{i\mathbf{k+q}}^{\rm W}| \varphi_{\mathbf{q}}^{\mathcal{S}}(\mathbf{r})| u_{j\mathbf{k}}^{\rm W} \rangle = \delta_{ij}  \\ 
 + i\mathbf{q} \cdot  \Big[  \mathbf{A}_{ij\mathbf{k}}^{\text{W}}  + \langle u_{i\mathbf{k}}^{\rm W}| \frac{V^{\textrm{Hxc},\boldsymbol{\mathcal{E}}}(\mathbf{r})}{e} | u_{j\mathbf{k}}^{\rm W} \rangle  \Big], 
\end{multline}
where $A_{ij\mathbf{k},\alpha}^{\text{W}} \equiv -i\left\langle \frac{\partial u_{i\mathbf{k}}^{\rm W}}{\partial k_\alpha}\middle| u_{j\mathbf{k}}^{\rm W} \right\rangle$ is the Berry connection.
The second term in \Eq{eq:long-range-new} has been found to be small~\cite{Brunin2020, Brunin2020a} and thus is neglected.
For this work, we extended the original  \textsc{EPW} implementation of WFPT~\cite{Lihm2021} to support quadrupoles and Berry connection contributions.

The momentum convergence using that approach is also presented in \Fig{fig:qconv} for diamond and shows the same type 
of accuracy as the variation observed between the two implementations of the perturbed potential in \textsc{Abinit} and \textsc{Quantum ESPRESSO}. 
This is actually remarkable as one would expect larger differences due to additional errors linked with the Wannierization quality itself, which is absent in the other two. 
In particular, the largest relative difference at convergence is 0.44\% for the band edges and 1.81\% for the CB, see Fig.~S4 of the SI~\cite{SI} for more details. 
In addition, we have performed a similar convergence analysis for BAs for the three cases using \textsc{Abinit}, \textsc{Quantum ESPRESSO}, and \textsc{EPW} and find a smooth linear convergence where the CB still oscillate by less than 3~meV for our largest 200$\times$200$\times$200 $\bq$-points grids.  
We show the results in Fig.~S5 of the SI~\cite{SI}. 
Notably, the AHC implementation inside \textsc{Abinit} and \textsc{Quantum ESPRESSO} gives similar results but the one using WFPT in \textsc{EPW} differs by up to 5~meV.
We will discuss this difference in more depth in the next subsection where we compare the different approaches.


Finally, we compute the ZPR of diamond using the adiabatic nonperturbative special displacement method (SDM) using the \textsc{ZG} software~\cite{Zacharias2020} which is a type of non-perturbative, supercell-based approach,
comparable to finite difference methods~\cite{Antonius2014,Monserrat2014,Monserrat2018} including the thermal lines approach~\cite{Monserrat2016,Monserrat2016a} and the use of non-diagonal supercells~\cite{LloydWilliams2015}.  
We stress that a perfect agreement on the ZPR between the SDM and the AHC or WFPT is impossible even under the adiabatic approximation because the SDM includes some anharmonic effects and the full Debye-Waller term, beyond the RIA.

The Williams-Lax theory~\cite{Williams1951,Lax1952} is a semiclassical approximation whereby the transition rate between a coupled electron-phonon state is evaluated using Fermi's golden rule.
The transition rate at a finite temperature is then obtained by performing a canonical thermal average.
One can then obtain the Kohn-Sham temperature-dependent electronic energies by averaging the position-dependent energies over thermal fluctuations~\cite{Zacharias2020}:
\begin{equation}\label{eq:obs}
\varepsilon_{n\bk}(T) = \prod_{\bq\nu} \int \frac{dz_{\bq\nu}}{\pi u_{\bq\nu}^2} e^{-|z_{\bq\nu}|^2/u_{\bq\nu}^2} \varepsilon_{n\bk}(\{ z_{\bq \nu} \}),
\end{equation}
where $z_{\bq\nu}$ are the normal mode coordinate, $u_{\bq\nu}^2 = (2n_{\bq\nu}(T) + 1 )(\hbar/2M_0 \omega_{\bq\nu})$ the mean-square displacement for the oscillator $\bq\nu$ with Bose-Einstein distribution $n_{\bq\nu}$.
The amplitudes in the normal mode coordinate $z_{\bq\nu}$ can be computed as a linear combination of atomic displacements $\Delta \boldsymbol{\tau}_{\kappa l}$ for atom $\kappa$ in the supercell with cell index $l$:
\begin{equation}
z_{\bq\nu} = N^{-1/2} \sum_{\kappa\alpha l} \sqrt{\frac{M_\kappa}{M_0}} e^{-i \bq \cdot \mathbf{R}_l} \mathbf{e}_{\kappa,\nu}^*(\bq) \Delta \boldsymbol{\tau}_{\kappa l} \,,
\end{equation}
and vice versa, where $N$ is the number of unit cells in the supercell.

The SDM is based on the realization that the number of configurational samples needed decreases with the size of the supercell.
In the limit of an infinite supercell size, a single configuration with atomic displacements is sufficient~\cite{Zacharias2016}. 
The configurational average in Eq.~\eqref{eq:obs} can be approximated by a single optimal configuration $\boldsymbol{\tau}_{\kappa l} = \boldsymbol{\tau}^0_{\kappa l} + \Delta \boldsymbol{\tau}^{\rm ZG}_{\kappa l}$
where $\boldsymbol{\tau}^0_{\kappa l}$ is the 0~K DFT ground-state atomic position and $\Delta \boldsymbol{\tau}^{\rm ZG}_{\kappa l}$ a displacement given by~\cite{Zacharias2020}
\begin{equation}
\Delta \boldsymbol{\tau}^{\rm ZG}_{\kappa l} = 2 \sqrt{\frac{M_0}{N M_\kappa}} \sum_{\bq \nu} S_{\bq\nu} u_{\bq\nu} \Re [ e^{i\bq \cdot \mathbf{R}_l} \mathbf{e}_{\kappa,\nu}(\bq)] \,,
\end{equation}
where $M_0$ is an arbitrary reference mass chosen to be the proton mass, $M_\kappa$ the mass of the atom $\kappa$, $ \mathbf{e}_{\kappa,\nu}(\bq)$ the phonon eigendisplacement vectors, and $S_{\bq\nu}$ are alternating signs ($\pm 1$) that optimize the approximation to Eq.~\eqref{eq:obs}.

Using Eq.~\eqref{eq:obs} and the SDM as implemented in the \textsc{ZG} code, we compute the adiabatic ZPR of diamond as a function of supercell size.
As seen in Fig.~\ref{fig:qconvZG}, for the CB, the convergence of \textsc{ZG} with respect to supercell-size is faster than that of the perturbative approaches with respect to the \textbf{q}-point grid; in the former we approach convergence with a 9$\times$9$\times$9 supercell,
corresponding to only a 9$\times$9$\times$9 \textbf{q}-point grid.
This faster convergence for states that are not band edges has been explained in Ref.~\onlinecite{Miglio2020} as a small contribution from linear terms in the case of the non-perturbative approach that smooths convergence. 
However, the cost is also larger as that supercell includes 1458 atoms which means 2930 bands, requiring a larger plane-wave energy cutoff of 80~Ry to describe them.  
With the largest supercell size, we obtain an adiabatic ZPR value of 136.5$\pm$4~meV, -284.6$\pm$10~meV, and -267.0$\pm$11~meV for the VBM, CB, and CBM, respectively.

\subsection{Comparison between the different codes and approaches}

\begin{figure}[t]
  \centering
  \includegraphics[width=0.99\linewidth]{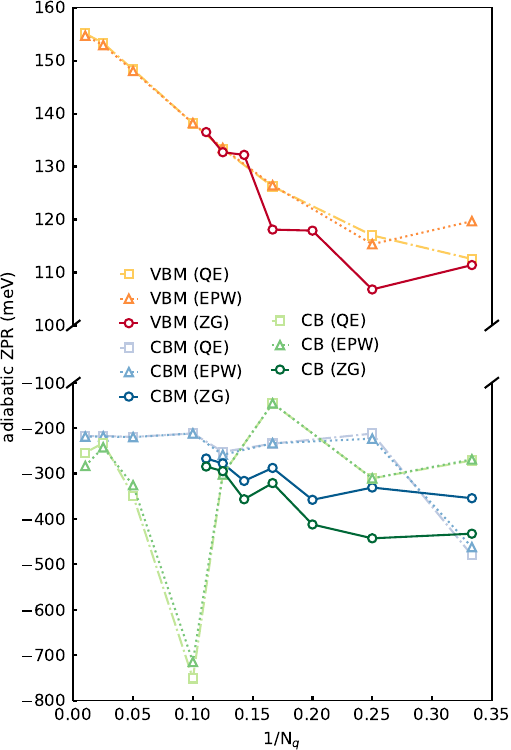}
  \caption{\label{fig:qconvZG}
\textbf{Comparison between methods for the adiabatic zero-point motion renormalization of diamond}.
Convergence of the valence band maximum (VBM), conduction band minimum (CBM) at 0.72$\Gamma X$ and conduction band at the zone center (CB) at 0~K using adiabatic Allen-Heine-Cardona calculation with the \textsc{Quantum ESPRESSO (QE)}, \textsc{EPW} codes, and finite displacement with the \textsc{ZG} code as a function of the number of atoms in the supercell. 
The largest supercell is a 9$\times$9$\times$9 one which includes 1458~atoms and corresponds to a 9$\times$9$\times$9 $\bq$-point grid. 
Due to the very large number of bands in the supercell calculations (up to 2930 bands), we increased the plane-wave energy cutoff to 80~Ry to describe them.
}
\end{figure}

We now assess the level of agreement between different codes implementing the same equations as well as the validity of certain approximations. 
Based on Figs.~\ref{fig:qconv} and \ref{fig:qconvZG}, and for the purpose of comparison, we consider a momentum grid integration of 100$\times$100$\times$100 $\bq$-grid, Sternheimer, 9 active bands, quadrupoles, and 5~meV smearing for the perturbative approaches and a 9$\times$9$\times$9 supercell size for the non-perturbative one.  
We assume the CBM to be located at 0.72$\Gamma$X in all cases.

We start by investigating the adiabatic ZPR and show the results for the four codes investigated in Table~\ref{table1}.
We find an excellent agreement between the \textsc{Abinit} and \textsc{Quantum ESPRESSO} implementation of the AHC theory for the VBM, CB, and CBM, with the largest absolute difference being 0.6~meV, therefore \emph{verifying} that plane-wave codes that use the same pseudopotential and implement the same theory do give the same result. 
When comparing the \textsc{EPW} implementation of WFPT to the AHC theory, we also find excellent agreement for the VBM and CBM with the largest difference being 0.7~meV. 
However, surprisingly, we find that WFPT overestimates the CB renormalization by 29~meV.
We further decompose the ZPR into its Fan and DW part and for each case, we also decompose the active and rest subspace contribution and report the results in Table~\ref{table2}.
For the CB case, we find that 2/3 of the difference between adiabatic WFPT and AHC theory comes from Fan term and 1/3 comes from the DW term in the rest subspace, the active DW part being the same.
We have not been able to understand the precise source of discrepancy in that case but, as we will soon see, the problem is not present in the case of the non-adiabatic solution.
Since in practice the non-adiabatic solution is always superior to the adiabatic one, we do not try to fix this and simply report that 
we could not \emph{validate} the adiabatic WFPT with the adiabatic AHC.

We then compared our adiabatic AHC result with the adiabatic special displacement methods.
For the VBM, the ZG approach underestimates the result by 18~meV but as can be seen on Fig.~\ref{fig:qconvZG}, the results are not fully converged. 
As demonstrated in Ref.~\onlinecite{Ponce2015}, the result can be linearly extrapolated to denser momentum grids. 
We therefore extrapolate the \textsc{ZG} results to 100$\times$100$\times$100 $\bq$-grid and obtain 158~meV, closer to the 155~meV of the AHC results.    
We infer that the remaining 2\% difference is mostly due to the RIA, as the anharmonicity is expected to be small in silicon.
We can perform a similar analysis for the CBM which gives an extrapolated value of -226~meV which is 4\% away from the AHC value.
We note that these values will change with larger ZG supercells and that a potential small error compensation with the anharmonic effect is possible.
As a result, we can confidently deduce that the effect of the RIA is below 10\% for diamond.
These findings are in line with the recent Ref.~\onlinecite{Mann2024} that reports a sub 5\% contribution from the RIA in diamond and silicon.

In addition, for the CB we obtain -284.6~meV with the ZG method, in quite good agreement with the AHC result of -254~meV.
However, as discussed previously and extensively in Ref.~\onlinecite{Ponce2015}, the CB is more challenging to converge due to energy resonances and we estimate that we are too far from convergence to be able to perform any meaningful extrapolation of the \textsc{ZG} ZPR. 
Overall, we note that the validity of the RIA was previously explored for specific $\bq$-points by direct comparison to finite difference in Ref.~\onlinecite{Ponce2014} but it is the first time that a quantitative estimate is given for the impact of RIA on the integrated ZPR.

We then compare the non-adiabatic results in Table~\ref{table1}.
In that case, the AHC as implemented in \textsc{Abinit} and \textsc{Quantum ESPRESSO} as well as WFPT all give excellent agreement. 
The largest absolute difference is 1.4~meV, occurring for the CBM.
Actually, looking at Table~\ref{table2} shows that the agreement is also excellent for the decomposed quantities, therefore both \emph{validating} and \emph{verifying} the theory and implementation of the non-adiabatic AHC in IR-inactive materials.

In the context of this work, we also mention the work of M.~Engel~\textit{et al.}~\cite{Engel2022} who computed the non-adiabatic ZPR of the indirect bandgap of diamond using the projector-augmented-wave (PAW) method~\cite{Blochl1994a} in the \textsc{VASP} software~\cite{Kresse1993,Kresse1996} with a treatment of the long-range part of the potential derivative.
The authors obtained 323~meV, in close agreement with our 330~meV reported in Table~\ref{table1}.

We then perform the same comparison on BAs to assess the effect of IR activity on the results. 
In that case, only the non-adiabatic ZPR can be computed, as the $\bq$ integral diverges in principle for the adiabatic ZPR~\cite{Ponce2015}.
In the case of the two AHC implementations, we verify that the agreement is excellent with a renormalization of the VBM at 0~K due
to the zero-point motion of 45.47$\pm$0.4~meV and -49.74$\pm$0.2~meV for the CBM, giving a sizable indirect bandgap reduction of 95~meV.    
We also find that the CB at the zone center has a renormalization of -73.65$\pm$0.2~meV which yields a direct bandgap reduction of 119~meV. 
Individual values and their decomposition can be found in Table~S1 of the SI~\cite{SI}.

Furthermore, we have also studied the impact of having semicore $d$ state explicitly included in the pseudopotential of arsenic and found very little impact except for a 5~meV increase in the CB ZPR (see Table~S2 of the SI~\cite{SI}). 
We have nonetheless used As with semicore state for all this work for definitiveness. 
We also note that in Table~S1 of the SI~\cite{SI} there is a different separation between active and rest space since the semicore states are part of the active space in the \textsc{Quantum ESPRESSO} calculation and part of the rest space in \textsc{Abinit}.
This, however, bears no consequences on the final observable.

Moreover, we looked at the difference between the WFPT and the AHC theory.
In the case of BAs, we do observe a larger difference with values of 43.37~meV, -66.60~meV, and -52.68~meV for the VBM, CB, and CBM, respectively. 
These results yield a maximal difference of 9.6\% between the WFPT and the average AHC values between the two codes.
Although reasonable, this is a significantly larger difference than the one observed for diamond.  
Looking at the Fan/DW decomposition, we find that the error made is similar on both terms and between 3~meV to 17~meV. 
Interestingly, in all cases, the errors compensate, and the resulting difference in the observable ZPR is smaller and ranges from 1.7~meV to 6.9~meV.
We therefore conclude that describing the quantum state with Wannier functions and their perturbation can yield errors of up to 10\% in polar materials.

Finally, we compare our result to Ref.~\cite{Bravi2019} where they compute the ZPR of BAs using a finite-displacement method and an 8$\times$8$\times$8 supercell.
Interestingly, since BAs has a weak LO-TO splitting due to highly covalent bond resulting in small Born effective charges~\cite{Hadjiev2014,Bravi2019}, see also Fig~S1 of the SI~\cite{SI}, adiabatic methods should remain quite valid~\cite{Miglio2020}.
For this reason, we also computed the ZPR of BAs using the adiabatic approximation with a 100$\times$100$\times$100 $\bq$-grid grid and obtained 50.2~meV, -47.7~meV, and -57.1~meV for the VBM, CB, and CBM, respectively.
This gives us an adiabatic ZPR reduction of the direct and indirect bandgap of 97.9~meV and 107.3~meV, respectively.    

In the quadratic approximation, which consists of neglecting terms beyond the second order the the eigenvalue expansion in terms of atomic displacement, they obtain a 97~meV reduction of the indirect bandgap of BAs due to the ZPR~\cite{Bravi2019}, in almost perfect agreement with our non-adiabatic 96.1~meV value from Table~\ref{table1} and 10\% smaller than our adiabatic value.
This agreement implies that the RIA should be quite good in BAs as well.
In that work, they further study the effect of anharmonicity via Monte Carlo integration and find a 10\% increase in the ZPR
as well as a further $\approx$ 10\% increase due to electron correlation using the hybrid Heyd-Scuseria-Ernzerhof functional (HSE)~\cite{Paier2006} instead of the PBE exchange-correlation functional.
They also find that both thermal expansion and spin-orbit-coupling (SOC) have negligible (within the statistical uncertainty of their stochastic calculations) effects on the ZPR~\cite{Bravi2019}.

\subsection{Impact of spin-orbit coupling}

The inclusion of spin-orbit-coupling (SOC) in the calculation impacts the electronic bandstructure and 
in particular, the VBM where the triple degeneracy is lifted resulting in spin-spit bands.
In the case of diamond, we computed the split-off energy to be 13.5~meV, and the corresponding non-adiabatic ZPR using WFPT to be 130.4~meV, -270.7~meV, and -199.4~meV for the VBM, CB, and CBM, respectively.  
By comparing these results with those without SOC in Table~\ref{table1}, we find that the effect of SOC 
on the ZPR is small and reduces the ZPR by up to 3\%. 
This is in line with previous studies~\cite{Bravi2019,BrousseauCouture2023} which shows that the effect of SOC is small and reduces the ZPR. 
The largest reported reduction is 30\% for CdTe which has a very large SOC and a split-off energy of 0.85~eV~\cite{BrousseauCouture2023}.

In contrast, the impact of SOC on the hole mobility is much larger and may lead to a 70\% increase even in materials with modest SOC~\cite{Ma2018,Ponce2018,Ponce2021}.
As the split band goes down in energy, the available scattering phase space is reduced, and the hole mobility is enhanced. 
Specifically in diamond, the room temperature Boltzmann transport equation hole mobility increases from 2265~cm$^2$/Vs to 2290~cm$^2$/Vs when including SOC~\cite{Ponce2021}.
Instead, for BAs, the room temperature hole mobility without SOC was reported to be 1387~cm$^2$/Vs~\cite{Bushick2020} and 2110~cm$^2$/Vs~\cite{Liu2018} when including SOC due to the large 216~meV spin-off splitting.
Indeed, the hole mobility with SOC is about 3/2 the one without due to the split-off bands going significantly down and not contributing to the resistivity.

\begin{figure}[t]
  \centering
  \includegraphics[width=0.99\linewidth]{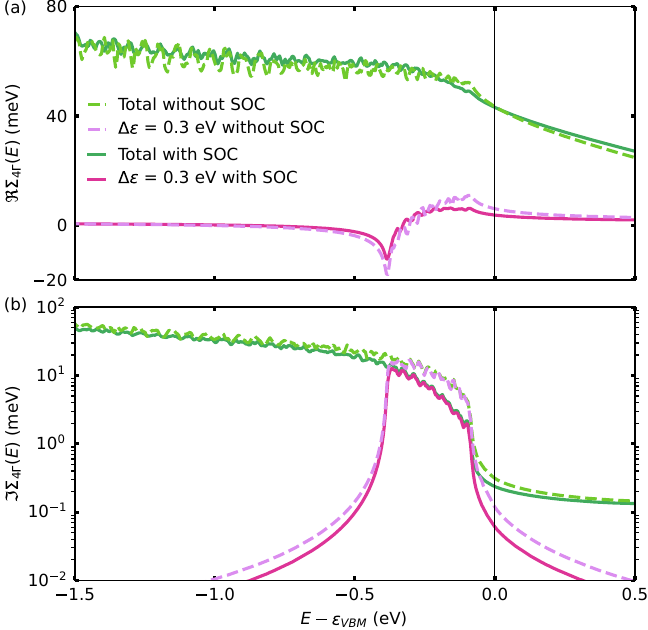} 
  \caption{\label{fig:selfenergy}
\textbf{Impact of spin-orbit coupling on the electron-phonon self-energy of BAs.}  
Real (a) and imaginary (b) part of the electron-phonon self-energy at the valence band maximum as a function of energy integrated on a 100$\times$100$\times$100 $\bq$-grid integration with 5~meV smearing. 
At the VBM, the real part is unaffected while the imaginary part is reduced by 23\% when including SOC. 
With pink/mauve lines we show the contribution to the self-energy from states $m\bk+\bq$ within 0.3~eV of the band edge.
}
\end{figure}

These two facts may appear contradictory as the carrier mobility in the self-energy relaxation time approximation (SERTA)~\cite{Ponce2020} is directly proportional to the inverse of the imaginary part of the Fan self-energy $\Sigma_{nn\bk}^{\rm Fan}(E)$ and the real and imaginary parts are related by the Kramers-Kronig relation.
Therefore, to study this aspect, we have computed in Fig.~\ref{fig:selfenergy} the energy-dependent real and imaginary parts of the self-energy of BAs for the VBM with and without SOC. 
Having established the agreement between codes, we here use \textsc{EPW} and WFPT for this analysis 
and use the same 100$\times$100$\times$100 $\bq$-grid integration with 5~meV smearing.  
With such a small smearing, there are visible oscillations that would require a denser $\bq$-grid to be reduced. 
However, the solution is smooth around the VBM energy which is what is used for the calculation of ZPR.

As pointed out by Ref.~\onlinecite{BrousseauCouture2023} and shown with green lines in Fig.~\ref{fig:selfenergy}, the relative reduction of the imaginary part of  $\Sigma_{nn\bk}(E)$ at the VBM due to SOC is much larger than the relative reduction of the real part.
At the VBM, the real part is unaffected while the imaginary part is reduced by 23\% when including SOC.
However away from the VBM, the real part shows changes upon including SOC, ensuring that the Kramers-Kronig relation is respected.    
To understand why the mobility is more affected by SOC, we also show with mauve lines in Fig.~\ref{fig:selfenergy} the $\Sigma_{nn\bk}(E)$ where we have only included in the phonon $\bq$ momentum integral the states for which the eigenvalues $\varepsilon_{m\bk+\bq}$ is within 0.3~eV of the VBM. 
Indeed, only the states close to the band edge will contribute to carrier mobility and 0.3~eV is a conservative energy window for the room temperature mobility~\cite{Ponce2021}.
In that case, the real part is reduced by 5\% at the VBM while the imaginary part is reduced by 48\% when including SOC. 
Since SOC impacts mostly the states close to the band edges, the mobility is impacted more strongly than the ZPR which has contributions from all the states. 
We have also performed the same analysis for diamond, see Fig.~S7 of the SI~\cite{SI}.
In that case, the impact of SOC is very small both for the real and imaginary parts of $\Sigma_{nn\bk}(E,T)$.

In conclusion, for states that are strongly impacted by SOC, it is expected that the impact on mobility will be larger than on ZPR as
the ZPR has contributions from all the states, while the mobility solely comes from a small energy region around the band edges.
This fact explains why the effect of SOC on the real and imaginary parts can be different, even though they are related by the Kramers-Kronig relation.
Given the small impact of SOC on the ZPR, we neglect it for the rest of this manuscript.


\subsection{Mass enhancement} \label{sec:mass}
The diagonal electron spectral function can be computed from the electron-phonon self-energy as:
\begin{equation}
A_{n\bk}(E,T) = -\frac{1}{\pi} \Im \frac{1}{E - \varepsilon_{n\bk} - \Sigma_{nn\bk}(E,T)}.
\end{equation} 

Due to the challenges in computing the DW term, many authors~\cite{Verdi2017,Riley2018,Caruso2018,Zhou2019,Chang2022,Abramovitch2023} (but not all~\cite{Nery2018,deAbreu2022}) have computed the spectral function by approximating the electron-phonon self-energy by the Fan-Migdal one $\Sigma_{nn\bk}^{\rm Fan}(E,T)$. 
The justification for such an approximation is that the DW term is static and might contribute mostly to a rigid shift of the eigenvalues. 
Therefore, we look at the $\bk$ momentum dependence of the DW term and the various approximations summarized in Fig.~\ref{fig:decomp} to see if such an approximation is sound.

One physical quantity associated with the momentum dependence is the mass enhancement due to electron-phonon coupling, which can be obtained from the effective mass of the renormalized band:
\begin{equation}
m_{n\alpha\beta}^{\rm ep}(T) = \hbar^2 \frac{\partial^2 E_{n\bk}^{\rm RS}(T)}{\partial k_\alpha \partial k_\beta} = m_{\alpha\beta}^* (1  + \lambda_{n\alpha\beta}(T)).
\end{equation}
Here, $E_{n\bk}^{\rm RS}(T)$ is the on-the-mass-shell renormalized energy from Eq.~\eqref{eq:rs} and $\lambda_{n\alpha\beta}(T)$ denotes the mass-enhancement factor.

We computed the DFT bare hole effective mass of diamond using \textsc{Quantum ESPRESSO} in the $X$ and $L$ direction for the light-hole and heavy-hole bands, $m^{\rm hh}_{\Gamma L}$,  $m^{\rm lh}_{\Gamma L}$,  $m^{\rm hh}_{\Gamma X}$, and  $m^{\rm lh}_{\Gamma X}$, as well as the longitudinal and transverse electron effective masses $m^{\rm e}_{\rm l}$ and $m^{\rm e}_{\rm t}$, respectively, and report them in Table~\ref{table3}.
We verified that we obtain the same (to 3 significant digits) DFT effective mass with \textsc{Abinit}.
We obtain longitudinal and transverse masses of 1.68~$m_0$ and 0.30~$m_0$, respectively, in relatively good agreement with the experimental values of 1.4~$m_0$ and 0.36~$m_0$~\cite{Nava1980}.
The hole effective masses show larger deviations since we neglect SOC.
However, due to Wannierization, the bare effective mass for \textsc{EPW} is slightly different and its value can be deduced from the mass enhancements $\lambda_{n\alpha\beta}$ reported in the table.

\begin{figure}[t]
  \centering
  \includegraphics[width=0.9\linewidth]{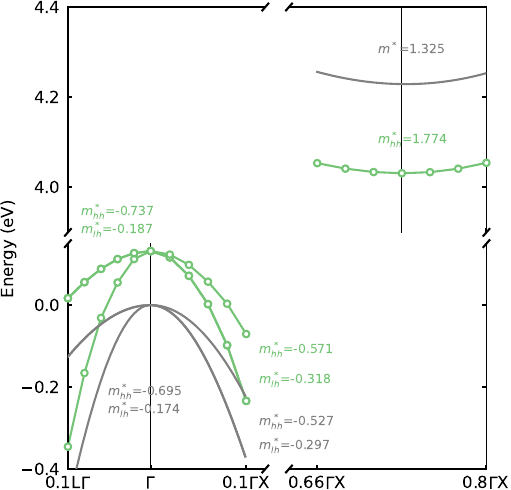} 
  \caption{\label{fig:mass}
\textbf{Renormalization of the bandstructure of diamond due to electron-phonon coupling.}  
Valence and conduction bands computed using DFT (gray), including the 0~K renormalization due to electron-phonon interaction (green). 
The effective masses along specific high-symmetry directions are indicated, providing the mass enhancement $m_{\alpha\beta}^{\rm ep} = m_{\alpha\beta}^*(1+\lambda_{\alpha\beta})$.
The calculations are made using the on-the-mass shell approximation of the non-adiabatic AHC implementation of \textsc{Quantum ESPRESSO} integrated on a 100$\times$100$\times$100 $\bq$-grid integration, including quadrupole, and 5~meV smearing.  
}
\end{figure}

We then compute the change of curvature due to electron-phonon interaction.
We find longitudinal and transverse effective masses of 1.89~$m_0$ and 0.33~$m_0$, respectively, with some small variation depending on the software, see Table~\ref{table3}. 
This corresponds to a mass enhancement of over 10\% in both directions.
As a representative case, we show in Fig.~\ref{fig:mass} the renormalization of the bandstructure of diamond using \textsc{Quantum ESPRESSO} and focusing on the band edges.

Interestingly, we note that the mass-enhancement due to electron-phonon coupling typically tends to increase the effective mass, often over-estimating the experiment. 
However, for diamond, $G_0W_0$ and quasiparticle self-consistent $GW$ (qs$GW$) both decrease the effective mass by about 10\%~\cite{Lfs2011}.
This effect is reminiscent of the overestimation of bandgaps using qs$GW$ which is then reduced closer to the experiment by the electron-phonon ZPR~\cite{Miglio2020}.

\begin{figure}[ht]
  \centering
  \includegraphics[width=0.9\linewidth]{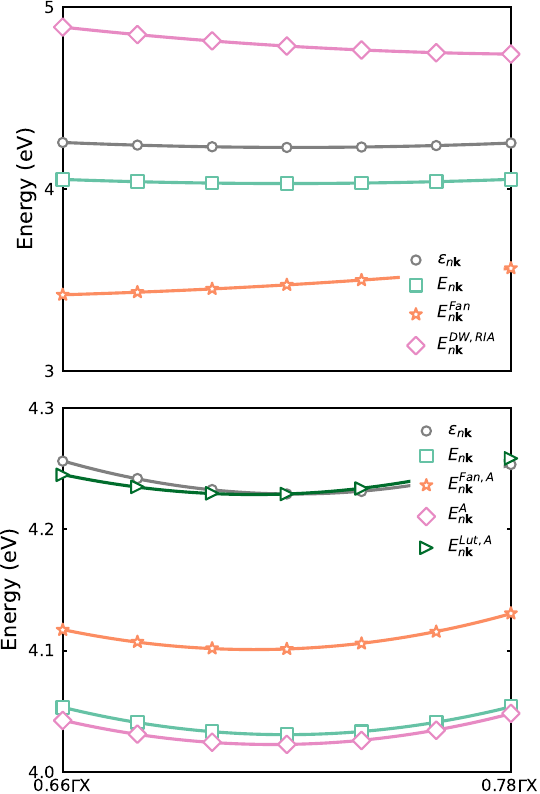} 
  \caption{\label{fig:mass2}
\textbf{Study of the conduction band minimum of diamond.}
The top panel presents a zoom-in around the CBM with DFT eigenvalues and including non-adiabatic ZPR at 0~K due to electron-phonon coupling ($E_{n\bk}$), the Fan term only ($E_{n\bk}^{\rm Fan}$), and the Debye-Waller (DW) term only ($E_{n\bk}^{\rm DW}$), respectively. 
Both the Fan and DW terms are large and of opposite magnitude and anisotropy.
Only the sum of both terms is observable.  
The bottom panel presents a further zoom-in showing three approximations using the active space (9 bands) only:
(i) the Fan term on the active space only $\Sigma_{n\bk}^{\rm Fan,A}$ using Eq.~\eqref{eq:approx_FanA}, (ii) active space only $\Sigma_{n\bk}^{\rm A}$ using Eq.~\eqref{eq:approx_A}, as well as (iii) the Luttinger approximation $\Sigma_{n\bk}^{\rm Lut, A}$ of Eq.~\eqref{eq:luttingerA}.  
The calculations are made using the on-the-mass shell approximation of the non-adiabatic AHC implementation of \textsc{Abinit} integrated on a 100$\times$100$\times$100 $\bq$-grid integration, including quadrupole, and 5~meV smearing. 
}
\end{figure}

We then compute the mass enhancement using the three active space approximations of Eqs.~\eqref{eq:luttingerA}-\eqref{eq:approx_FanA}.
We first show in Fig.~\ref{fig:mass2} the renormalization of the conduction band of diamond and find that the 
Fan and DW terms taken separately are very large and of opposite curvature.
Therefore, the corresponding mass enhancement using Eq.~\eqref{eq:approx_Fan_only}, which neglects the DW term entirely, is completely off. 
Instead, if we use Eq.~\eqref{eq:luttingerA} or Eq.~\eqref{eq:approx_FanA}, applying the Luttinger or Fan only approximation on the active space contribution and neglecting the rest-space contribution, we obtain an 
effective mass that is on average 9.5\% overestimated with respect to the WFPT reference.
We note that $\Sigma_{n\bk}^{\rm Fan,A}$ and $\Sigma_{n\bk}^{\rm Lut, A}$ give the same mass enhancement by construction but a different absolute value. 
As seen in the lower panel of Fig.~\ref{fig:mass2}, $\Sigma_{n\bk}^{\rm Fan,A}$ gives an underestimated ZPR that depends significantly on the size of the active space. 
The Luttinger approximation $\Sigma_{n\bk}^{\rm Lut, A}$, which removes the ZPR by construction, might therefore be preferred to avoid confusion.

Interestingly, computing the Fan and DW terms on the active space using Eq.~\eqref{eq:approx_A} reduced the difference with respect to the WFPT to 5.8\%.     
It also gives an excellent ZPR (4\% overestimation) that does not strongly depend on the active space. 
This last option is therefore the recommended approximation when working in a Wannier framework without a WFPT implementation.
We note that the main drawback is that the mass-enhancement then depends on the number of states included in the active space. 
We therefore recommend using the smallest active space that includes the full description of the bands for which the mass enhancement is computed.

\subsection{Spectral functions} \label{sec:spectral}

We conclude this study by taking a brief look at the spectral function computed using the \textsc{Abinit} and \textsc{EPW} codes
to verify the accuracy of that physical property. 
Indeed, the spectral function is a more sensitive quantity than the ZPR as it includes the full description of the energy-dependent electron-phonon self-energies.
We compute the diagonal spectral function from the self-energy as 
\begin{multline}
A_{n\bk}(E,T) = \\
-\frac{1}{\pi}\frac{\Im \Sigma_{n\bk}(E,T)}{(E-\varepsilon_{n\bk} - \Re \Sigma_{n\bk}(E,T))^2 + (\Im \Sigma_{n\bk}(E,T))^2}.
\end{multline}

\begin{figure*}[t]
  \centering
  \includegraphics[width=0.99\linewidth]{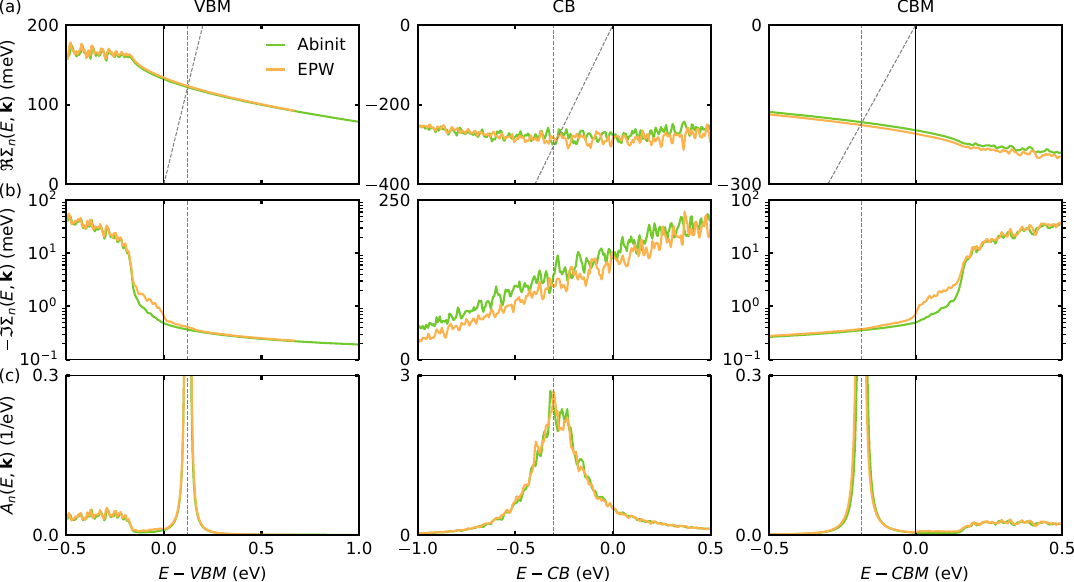} 
  \caption{\label{fig:spectral}
\textbf{Electron self-energy and spectral function of diamond due to electron-phonon coupling.}  
(a) Real and (b) imaginary part of the electron self-energy along with (c) the corresponding spectral function, computed with \textsc{Abinit} and \textsc{EPW} using WFPT at 300~K. 
The calculations are made using a 100$\times$100$\times$100 $\bq$-grid integration and 5~meV smearing for the calculation of the self-energies. 
}
\end{figure*}

We present in Fig.~\ref{fig:spectral} the spectral function for the VBM, CB, and CBM states around their respective bare energy eigenvalues.
We show with dashed lines in the top panels the geometrical solution of Eq.~\eqref{eq:dm} which is obtained as the intersection of the 
$E-\varepsilon_{n\bk}$ line and $\Re \Sigma_{n\bk}(E, T)$, where $\varepsilon_{n\bk}$ is the VBM, CB, or CBM depending on the case. 
For diamond we obtain 119.7~meV, -304.8~meV, and -185.0~meV for the VBM, CB, and CBM, respectively.    
These values are comparable to the RS approximation $E_{n\bk}^{\rm RS}(T)$ of Eq.~\eqref{eq:rs} values of 131.7~meV, -279.3~meV, and -199.2~meV for the VBM, CB, and CBM, respectively.   

Interestingly, overall the results are in excellent agreement between the two codes with only small differences observed in the imaginary part of the self-energy. 
The small oscillations seen away from the band edges are characteristic and due to numerical resonances when states at $\bk$ and $\bk+\bq + \omega_{\bq}$ have similar energies. 
These oscillations can be systematically reduced by using denser momentum grids but we restrict ourselves to 100$\times$100$\times$100 $\bq$-grid to be consistent with most of the results presented in this work. 

\section{DISCUSSION}\label{sec:conclusion}

In this work, we have successfully \emph{verified} four first-principles codes and \emph{validated} three methods to compute zero-point renormalization of the bandgap, mass enhancement, and spectral functions, which allowed us to fix bugs and improve the functionalities of the software as well as giving more insights into the accuracy and limitation of the different methods.  
We report for the first time the slow sum-over-state convergence toward the Sternheimer-based results in the case of solids and explain why certain electronic states converge faster than others.
We show that the rigid-ion approximation is a good approximation in solids. 
We explained the small impact of spin-orbit coupling on the real part of the electron-phonon self-energy and the potential large impact on its imaginary part. 
We find that the Debye-Waller term is momentum-dependent and should be included, at least in the active subspace, to obtain accurate mass enhancement. 
We hope that such an effort will stimulate a global effort from the community to systemically verify and validate existing and new codes and methods. 

\section{METHODS}\label{sec:method}

\subsection{DFT calculations}
If not stated otherwise, all Brillouin-Zone integrals use a zone-centered homogeneous 8$\times$8$\times$8 $\bk$-point grids for electrons and 8$\times$8$\times$8 $\bq$-point grids for phonons.   
For diamond and BAs, we use a plane-wave energy cutoff of 40~Ha.
The DFT and DFPT self-consistent cycles are converged to tight values of 10$^{-20}$ Ry$^2$/$e^2$ or smaller.

\subsection{Optimization of EPW}

To improve performance, we implemented an optimized Fourier transform whereby a 3D Fourier transform is performed as a 1D nested Fourier transform along the $x$, $y$, and $z$ directions, respectively. 
This optimized Fourier transform is the same as Eqs.~(8)-(9) of Ref.~\onlinecite{Kaye2023} and allows for a cost reduction for $\bq$ points on a uniform mesh.
In addition, we rewrote the Fourier interpolation using Wigner-Seitz lattice vectors following the \textsc{WannierBerri} code~\cite{Tsirkin2021} [see Eqs.~(21)-(23) of Ref.~\onlinecite{Tsirkin2021}].
This refactoring enables an efficient call to \textsc{LAPACK} routines instead of a \emph{for} loop.
Both optimizations were developed and used in Ref.~\onlinecite{Lihm2021} and recently added into \textsc{EPW}.

\subsection{Software improvements resulting from this verification and validation effort}

For the \textsc{Abinit} code, we have implemented the support for adiabatic AHC band renormalization when using Fourier interpolation of the potential.
We introduced a new input variable \texttt{eph\_ahc\_type} for this.

For \textsc{Quantum ESPRESSO}, we have implemented support for adiabatic AHC band renormalization when using Fourier interpolation of the potential via the input \texttt{adiabatic = .true.}.
We have also implemented quadrupole support for the long-range part of the perturbed potential for Fourier interpolation.
The support for decomposing the Fan and Debye-Waller terms inside the active space was also added.
Also, we enabled the use of symmetry and the irreducible $\bq$ points for the band renormalization calculation.

For the WFPT, the original implementation of Ref.~\cite{Lihm2021} was merged to the \textsc{EPW}~\cite{Lee2023} code 
(available since v5.8).
Additional support for quadrupole, 2D~\cite{Ponce2023}, and Berry connection~\cite{Ponce2023a} was added. 
We have fixed a bug in WFPT related to the $\bq$ vector representation in the addition of the long-range electron-phonon coupling (fixed in EPW v5.9).
We have also fixed a bug in \textsc{EPW} related to the calculation of the Berry connection (fixed in EPW v5.9).

\section{Data availability}
The underlying code, pseudopotentials, all inputs and outputs, and
the scripts to produce the figures for this work are available on the Materials Cloud Archive for complete reproducibility and can be accessed with DOI:\href{10.24435/materialscloud:xr-ce}{10.24435/materialscloud:xr-ce}.

\section{Code availability}
All the codes used for this work are free and open-source. 

\section{acknowledgments}
The authors would like to thank Matteo Giantomassi and Xavier Gonze for useful discussions. 
S.P. and J.-M.L. acknowledge support from the Fonds de la Recherche Scientifique de Belgique (FRS-FNRS). 
This work was supported by the Fonds de la Recherche Scientifique - FNRS under Grants number T.0183.23 (PDR) and  T.W011.23 (PDR-WEAVE). 
This publication was supported by the Walloon Region in the strategic axe FRFS-WEL-T.
J.-M.L. and C.-H.P. acknowledge support from Korean-NRF No-2023R1A2C1007297.
Computational resources have been provided by the PRACE award granting access to MareNostrum4 at Barcelona Supercomputing Center (BSC), Spain and Discoverer in SofiaTech, Bulgaria (OptoSpin project id. 2020225411), and by KISTI (KSC-2022-CRE-0407), and by the Consortium des Équipements de Calcul Intensif (CÉCI), funded by the FRS-FNRS under Grant No. 2.5020.11 and by the Walloon Region, as well as computational resources awarded on the Belgian share of the EuroHPC LUMI supercomputer.

\section{Author contributions}
S.P. designed the project, performed the calculations and wrote the first draft of the manuscript. 
J.-M.L. analyzed the results, revised and implemented new functionalities to the WFPT and QE AHC codes, and contributed to writing the manuscript.
J.-M.L. and C.-H.P. provided early access to the WFPT code.
S.P. and C.-H.P. supervised the project.
All authors proofread and approved the final manuscript.

\section{Competing interests}
The authors declare no competing interests.

\section{Additional information}
The online version contains supplementary materials available at [TBA by the Editor]. 

Correspondence and requests for materials should be addressed to Samuel Ponc\'e.

\bibliography{Bibliography}

\clearpage

\begin{table}[t]
  \caption{\label{table1}
Non-adiabatic and adiabatic zero-point renormalization (ZPR) of the valence band maximum (VBM), conduction band at the zone-center (CB), and conduction band minimum (CBM)
for diamond using the Allen-Heine-Cardona (AHC) theory as implemented in the \textsc{Abinit} (AB) and \textsc{Quantum ESPRESSO} (QE) codes, as well as using the Wannier function perturbation theory (WFPT) within the \textsc{EPW} code. 
In all cases, we use Sternheimer, 8 active bands, quadrupoles, a 100$\times$100$\times$100 $\bq$-grid integration, and a 5~meV smearing. 
The adiabatic ZPR is also shown using the special displacement method as implemented in the \textsc{ZG} code.   
}
\begin{tabular}{ c c c c c}
  \toprule\\
 \multicolumn{5}{c}{diamond adiabatic ZPR (meV)}   \\
             & AB AHC & QE AHC & WFPT    & ZG  \\ 
\hline             
VBM          &  \,\,154.5 &  \,\,155.1  &  \,\,154.7 & \,\,136.5   \\
CB           &     -254.5 &     -254.2  &     -283.5 &    -284.6  \\
CBM          &     -218.0 &     -217.7  &     -218.7 &    -267.0   \\
indirect gap & \,\,372.5  &  \,\,372.8  &  \,\,373.4 & \,\,403.5  \\
  direct gap & \,\,409.0  &  \,\,409.3  &  \,\,438.2 & \,\,421.1  \\    
 & & & & \\    
 \multicolumn{5}{c}{diamond non-adiabatic ZPR (meV)}   \\
             & AB AHC & QE AHC & WFPT   \\
\hline
VBM          &  \,\,131.9 &  \,\,131.8 &  \,\,131.7  \\
CB           & -278.8 & -278.4 & -279.3   \\
CBM          & -198.3 & -197.8 & -199.2  \\
indirect gap &  \,\,330.2 &  \,\,329.6 &  \,\,331.0   \\
  direct gap &  \,\,410.7 &  \,\,410.1 &  \,\,411.1   \\     
 & & & & \\    
 \multicolumn{5}{c}{BAs non-adiabatic ZPR (meV)}   \\
             & AB AHC & QE AHC & WFPT   \\
\hline
VBM          &  \,\,45.1 &  \,\,45.9 &  \,\,43.4  \\
CB           & -73.5 & -73.8 & -66.6   \\
CBM          & -49.6 & -49.9 & -52.7  \\
indirect gap &  \,\,94.6 &  \,\,95.8 &  \,\,96.1  \\
  direct gap &  118.5 &  119.7 &  110.0   \\     
  \botrule
\end{tabular}  
\end{table}

\begin{table}[t]
  \caption{\label{table2}
Non-adiabatic and adiabatic zero-point renormalization (ZPR) decomposition into the Fan and Debye-Waller in the rigid-ion approximation (DW) terms of the valence band maximum (VBM), conduction band at the zone-center (CB), and conduction band minimum (CBM)
for diamond using the Allen-Heine-Cardona (AHC) theory as implemented in the \textsc{Abinit} (AB) and \textsc{Quantum ESPRESSO} (QE) codes, as well as using the Wannier function perturbation theory (WFPT) within the \textsc{EPW} code. 
In all cases, we use Sternheimer, 8 active bands, quadrupoles, a 100$\times$100$\times$100 $\bq$-grid integration, and a 5~meV smearing.  
}
\begin{tabular}{ r r r r r r r}
  \toprule\\
& \multicolumn{6}{c}{adiabatic ZPR (meV)}   \\
             & \multicolumn{2}{c}{AB AHC} & \multicolumn{2}{c}{QE AHC} & \multicolumn{2}{c}{WFPT}    \\
             & Fan & DW & Fan & DW & Fan & DW \\    
\hline             
        
VBM - active &    96.6 &   77.7 &  97.9   &   77.4 &    97.3 &   77.7 \\
    - rest   & -1673.3 & 1653.6 & -1672.7 & 1652.5 & -1663.5 & 1643.2 \\ 
    - full   & -1576.8 & 1731.3 & -1574.8 & 1729.9 & -1566.2 & 1720.9 \\
 CB - active &  -267.1 &  -30.7 &  -266.3 &  -30.6 &  -295.7 &  -30.7 \\
    - rest   & -1580.6 & 1622.2 & -1580.2 & 1622.9 & -1570.9 & 1613.8 \\    
    - full   & -1847.6 & 1593.1 & -1846.5 & 1592.3 & -1866.6 & 1583.1 \\
CBM - active &  -147.4 &  -78.6 &  -147.5 &  -78.3 &  -148.1 &  -79.1 \\
    - rest   &  -627.0 &  635.1 &  -627.4 &  635.5 &  -619.4 &  627.9 \\    
    - full   &  -774.5 &  556.5 & -774.9  & 557.2  & -767.5 &  548.8 \\
 & & & & \\  
 & \multicolumn{6}{c}{non-adiabatic ZPR (meV)}   \\
             & \multicolumn{2}{c}{AB AHC} & \multicolumn{2}{c}{QE AHC} & \multicolumn{2}{c}{WFPT}    \\
             &     Fan &    DW & Fan & DW & Fan & DW \\       
\hline
VBM - active &    74.0 &   77.7 &  74.5   &   77.4 &    74.4 &   77.7 \\
    - rest   & -1673.3 & 1653.6 & -1672.7 & 1652.5 & -1663.5 & 1643.2 \\
    - full   & -1599.4 & 1731.3 & -1598.1 & 1729.9 & -1589.1 & 1720.9 \\
CB  - active &  -291.3 &  -30.7 &  -290.5 & -30.6  &  -291.5 &  -30.7 \\
    - rest   & -1580.6 & 1622.2 & -1580.2 & 1622.9 & -1570.9 & 1613.8 \\    
    - full   & -1871.8 & 1593.1 & -1870.7 & 1592.3 & -1862.4 & 1583.1 \\
CBM - active &  -127.7 &  -78.6 &  -127.7 &  -78.3 &  -128.6 &  -79.1 \\
    - rest   &  -627.0 &  635.1 &  -627.4 &  635.5 &  -619.4 &  627.9 \\     
    - full   &  \,\,\,-754.7 & \,\,\,556.5 &  \,\,\,-755.0 &   \,\,\,557.2 & -748.0 & 548.8 \\
  \botrule
\end{tabular}  
\end{table}

\begin{table}[b]
  \caption{\label{table3}
Non-adiabatic effective masses (in the unit of electron mass in vacuum) of diamond at 0~K due to electron-phonon coupling using the AHC theory as implemented in the \textsc{Abinit} (AB) and \textsc{Quantum ESPRESSO} (QE) codes
as well as the WFPT, active space, and Luttinger (Lut) approximations within the \textsc{EPW} code. 
By construction, the active Fan only and active Luttinger have the same mass enhancement ($\lambda$) but different absolute values. 
In all cases, we use 9 active bands and quadrupoles. 
}
\begin{tabular}{ l r r r r r r}
  \toprule
                             & DFT    & AHC AB     & AHC QE      & WFPT &  Exp. &   \\
\hline
$m_{\rm l}^{\rm e}$          &  1.677 &  1.892 &   1.897 &  1.737  &  1.4~\cite{Nava1980}  & \\ 
$m_{\rm t}^{\rm e}$          &  0.300 &  0.336 &   0.360 &  0.323  & 0.36~\cite{Nava1980}  & \\ 
$m_{\Gamma \rm L}^{\rm hh}$  & -0.729 & -0.824 &  -0.751 & -0.811  & 2.12~\cite{Rauch1962} & \\  
$m_{\Gamma \rm L}^{\rm lh}$  & -0.192 & -0.181 &  -0.214 & -0.177  &  0.7~\cite{Rauch1962} & \\ 
$m_{\Gamma \rm X}^{\rm hh}$  & -0.535 & -0.603 &  -0.584 & -0.661  & 2.12~\cite{Rauch1962} & \\  
$m_{\Gamma \rm X}^{\rm lh}$  & -0.339 & -0.325 &  -0.355 & -0.333  &  0.7~\cite{Rauch1962} & \\ 
& & & & \\
\hline
                            &   bare &                                   &   WFPT & \multicolumn{3}{c}{Active space approximation}  \\
                            &    EPW &                                   &        &    Fan & Fan+DW & Lut \\
$m_{\rm l}^{\rm e}$         &  1.546 &  $\lambda_{\rm l}^{\rm e}$        &  0.123 &  0.136 &  0.125 &  0.136 \\ 
$m_{\rm t}^{\rm e}$         &  0.293 &  $\lambda_{\rm t}^{\rm e}$        &  0.103 &  0.117 &  0.099 &  0.117 \\ 
$m_{\Gamma \rm L}^{\rm hh}$ & -0.715 & $\lambda_{\Gamma \rm L}^{\rm hh}$ &  0.134 &  0.140 &  0.131 &  0.140 \\ 
$m_{\Gamma \rm L}^{\rm lh}$ & -0.192 & $\lambda_{\Gamma \rm L}^{\rm lh}$ & -0.077 & -0.069 & -0.081 & -0.069 \\     
$m_{\Gamma \rm X}^{\rm hh}$ & -0.552 & $\lambda_{\Gamma \rm X}^{\rm hh}$ &  0.198 &  0.215 &  0.167 &  0.215 \\  
$m_{\Gamma \rm X}^{\rm lh}$ & -0.334 & $\lambda_{\Gamma \rm X}^{\rm lh}$ & -0.001 &  0.006 &  0.001 &  0.006 \\  
  \botrule
\end{tabular}  
\end{table}

\end{document}